\documentclass[sigconf]{acmart}
\AtBeginDocument{%
  }

\setcopyright{acmlicensed}
\copyrightyear{2018}
\acmYear{2018}
\acmDOI{XXXXXXX.XXXXXXX}
\acmConference[Conference acronym 'XX]{Make sure to enter the correct
  conference title from your rights confirmation email}{June 03--05,
  2018}{Woodstock, NY}
\acmISBN{978-1-4503-XXXX-X/2018/06}

\usepackage{lipsum}
\usepackage{enumitem}
\usepackage{amsmath}
\usepackage{multicol,multirow,array}
\usepackage{subcaption,adjustbox}
\usepackage[dvipsnames]{xcolor}
\newcommand{\ToolX}{MFLI}

\usepackage{subcaption}
\begin{document}

\title{Rethinking ANN-based Retrieval: Multifaceted Learnable Index for Large-scale Recommendation System}


\author{Jiang Zhang, Yubo Wang, Wei Chang, Lu Han, Xingying Cheng, Feng Zhang, Min Li, Songhao Jiang, Wei Zheng, Harry Tran, Zhen Wang, Lei Chen, Yueming Wang, Benyu Zhang, Xiangjun Fan, Bi Xue, Qifan Wang}
\email{{jiangzhang2024,yubowang,wqfcr}@meta.com}
\affiliation{%
  \institution{Meta Platforms, Inc.}
  \state{}
   \country{}
}

\renewcommand{\shortauthors}{Jiang Zhang et al.}

\begin{abstract}

Approximate nearest neighbor (ANN) search is widely used in the retrieval stage of large-scale recommendation systems. In this stage, candidate items are indexed using their learned embedding vectors, and ANN search is executed for each user (or item) query to retrieve a set of relevant items. However, ANN-based retrieval has two key limitations. First, item embeddings and their indices are typically learned in separate stages: indexing is often performed offline after embeddings are trained, which can yield suboptimal retrieval quality—especially for newly created items. Second, although ANN offers sublinear query time, it must still be run for every request, incurring substantial computation cost at industry scale.
In this paper, we propose \textbf{M}ulti\textbf{F}aceted \textbf{L}earnable \textbf{I}ndex (\ToolX), a scalable, real-time retrieval paradigm that learns multifaceted item embeddings and indices within a unified framework and eliminates ANN search at serving time. Specifically, we construct a multifaceted hierarchical codebook via residual quantization of item embeddings and co-train the codebook with the embeddings. We further introduce an efficient multifaceted indexing structure and mechanisms that support real-time updates. At serving time, the learned hierarchical indices are used directly to identify relevant items, avoiding ANN search altogether.
Extensive experiments on real-world data with billions of users 
show that \ToolX~ improves recall on engagement tasks by up to 11.8\%, cold-content delivery by up to 57.29\%, and semantic relevance by 13.5\% compared with prior state-of-the-art methods. We also deploy \ToolX~ in the system and report online experimental results demonstrating improved engagement, less popularity bias, and higher serving efficiency.

\end{abstract}




\maketitle

\section{Introduction}
\label{sec:intro}
\begin{figure}
    \centering
    \includegraphics[width=0.95\linewidth]{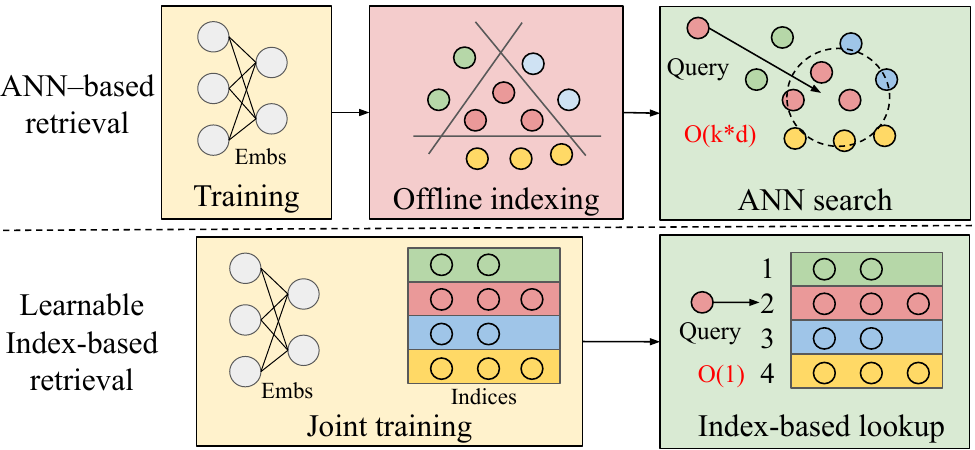}
    \vspace{-2mm}
    \caption{Traditional ANN-based retrieval contains three stages: item embedding learning, offline item indexing, and online ANN search at serving time. In contrast, LI-based retrieval jointly learns item embeddings and indices during training, and performs fast index lookup during serving.}
    \label{fig:intro}
    \vspace{-5mm}
\end{figure}

%
Approximate nearest neighbor (ANN) search \cite{aumuller2023recent,halder2024enhancing} lies at the core of large-scale retrieval in recommendation systems \cite{naumov2019deep,ko2022survey,simhadri2022results}, particularly when the candidate set reaches billions of items and exhaustive scanning becomes computationally infeasible \cite{huang2025comprehensive}. In general, ANN approaches construct an index over the item corpus such that similar items are mapped into the same (or adjacent) regions of the index, enabling efficient candidate item retrieval.

Despite their broad adoption, traditional ANN methods face two major challenges.
First, item indexing is typically performed offline and is decoupled from item-embedding learning. Because indexing is refreshed only periodically while the model is updated continuously online, the resulting indices can become stale and inconsistent across refresh cycles, which limits their effectiveness \cite{yan2024trinity,rajput2023recommender}. Moreover, constructing and refreshing indices in real time is expensive due to the heavy offline pipeline. This is particularly problematic for large-scale video recommendation, where millions of new videos may be created daily and freshness is important \cite{harichandan2025comprehensive}.
Second, the serving cost is high and often increases with candidate set size. For example, the widely used approach, the Inverted File Index (IVF) \cite{johnson2019billion}, probes a subset of nearby indices and then computes the k-nearest neighbors within those indices. At billion-scale, however, this still requires substantial online computation. While graph- and tree-based methods \cite{malkov2018efficient,zhu2018learning} can improve search speed, they typically incur significant memory overhead and are difficult to parallelize efficiently on modern GPU accelerators \cite{xue2025silvertorch}, limiting their serving efficiency (see Figure~\ref{fig:intro}). Therefore, it is an important research problem to address these effectiveness and efficiency limitations in traditional ANN-based retrieval approaches. 
Recent work on learning item indexing structures constructs item indices during training using Vector Quantization (VQ) techniques \cite{yan2024trinity,bin2025real}, aiming to mitigate inconsistencies between item-embedding learning and indexing. However, these approaches typically rely on a single index structure, which can limit representational capacity and reduce the diversity of retrieved items. Moreover, these works often underemphasize major system-level challenges of real-time index updates, scalability, and flexibility in index design.
To address these gaps, we propose \textbf{M}ulti-\textbf{F}aceted \textbf{L}earnable \textbf{I}ndex (MFLI), a scalable, real-time retrieval framework that jointly learns multifaceted item indices alongside item embeddings. We further describe MFLI’s system design for industrial-scale recommendation systems, covering both modeling and systems considerations needed to enable robust, real-time, multi-way item retrieval. 
At the high level, \ToolX~ comprises two key components (Figure~\ref{fig:intro}): 1) \textit{Joint item embedding and index training}. We introduce a unified training module that jointly learns item embeddings and a multifaceted, multi-layer codebook structure. Using distinct objective functions, the module produces richer and more diverse item representations, and enables each item to be mapped to multiple sub-codebooks in parallel.
2) \textit{Efficient serving via direct multifaceted lookup}. At serving time, the learned indices are used for retrieval through direct multifaceted item lookup, eliminating the traditional ANN search process and enabling efficient candidate retrieval. The retrieved candidates are then passed to a reranker, which performs per-index reranking to produce the final ranked list.

Building such systems presents several key challenges. First, maintaining balanced index (cluster) sizes during online training to prevent indices from becoming too large or too small. We apply a set of index-balancing strategies during training and introduce a split-and-merge procedure that automatically splits oversized indices and prunes undersized ones at serving time (Section \ref{subsec:train}). Second, designing an indexing scheme that enables efficient storage and retrieval of multifaceted item-to-index and index-to-item mappings. We propose a unified representation for multifaceted indices, together with a compact data structure that supports efficient lookup (Section \ref{subsec:publish}). Third, updating the indexing structure in real time, since item embeddings are refreshed asynchronously at different cadences during training and some candidate items may never appear in the training data. We develop an efficient delta-update strategy: a main stream periodically synchronizes indices for all items in the candidate pools, while a fast stream computes indices for fresh items in real time. At serving time, once an index is selected, we prefetch both existing and fresh items associated with that index in parallel and then merge the results, enabling efficient and effective index updates (Section \ref{subsec:serving}).

We conduct extensive evaluations of \ToolX~ on real-world data from a commercial platform serving billions of users.
\ToolX~ improves recall on engagement tasks by up to 11.8\%, recall for cold-content delivery by up to 57.29\%, and item semantic relevance by 13.5\% over prior SOTAs (Section \ref{subsec:main}). We further deploy \ToolX~ in the system and report online experimental results showing improved engagement, reduced popularity bias, and increased serving efficiency (Section \ref{subsec:online}). Last, we present comprehensive analysis of the learnt index distribution (Section \ref{subsec:analysis}). Our key contributions are summarized as follows:
\begin{itemize}[leftmargin=*]
  \item We propose \ToolX, a scalable real-time retrieval framework that jointly learns multifaceted item indices alongside item embeddings and eliminates the ANN process during serving.
  \item We introduce a split-and-merge procedure with guaranteed index balance, a unified and compact multifaceted-index storage for fast lookup, and a real-time delta-update pipeline that keeps the indexing structure fresh, addressing critical system challenges.
  \item We evaluate \ToolX~ on real-world data
  and show superior performance over prior SOTAs. The \ToolX\ is deployed in recommendation system, with online experiments validating its effectiveness.
\end{itemize}
\section{Related Work}
\noindent \textbf{ANN-based Retrieval.} Approximate nearest neighbor (ANN) search has been widely adopted in large-scale retrieval systems \cite{aumuller2023recent}.
Early work on ANN search proposed a variety of Euclidean-distance-based indexing methods, including k-means-based clustering \cite{johnson2019billion, douze2025faiss} and product quantization \cite{jegou2010product}. To improve retrieval accuracy, later approaches explored hashing-based methods \cite{shrivastava2014asymmetric,andoni2015practical,spring2017new}, graph-based methods \cite{malkov2014approximate,malkov2018efficient,chen2022approximate}, as well as tree-based methods \cite{houle2014rank,muja2014scalable,hyvonen2016fast}. However, these methods typically construct indices offline and are not well suited to GPU parallelization, causing mismatches between the item embeddings and reducing timeliness and scalability of large-scale online recommendation systems.

\noindent \textbf{Index Learning.} To overcome the limitations in ANN-based retrieval, recent work has proposed online indexing methods (e.g. Tree-based index models \cite{zhu2019joint,zhu2018learning}, DeepRetrieval \cite{gao2020deep}, Trinity \cite{yan2024trinity},   Streaming VQ \cite{bin2025real}, MERGE \cite{yan2026merge}) that learn item indices during training, mitigating the inconsistency between item embedding and index, and reducing serving latency and memory footprint. However, these approaches typically depend on a single index structure, which can constrain representational capacity and limit the diversity of retrieved items. Moreover, they often underemphasize important system-level challenges, including real-time index updates, scalability, and flexibility, in index design.
\begin{figure*}[t]
    \centering
    \includegraphics[width=0.93\linewidth]{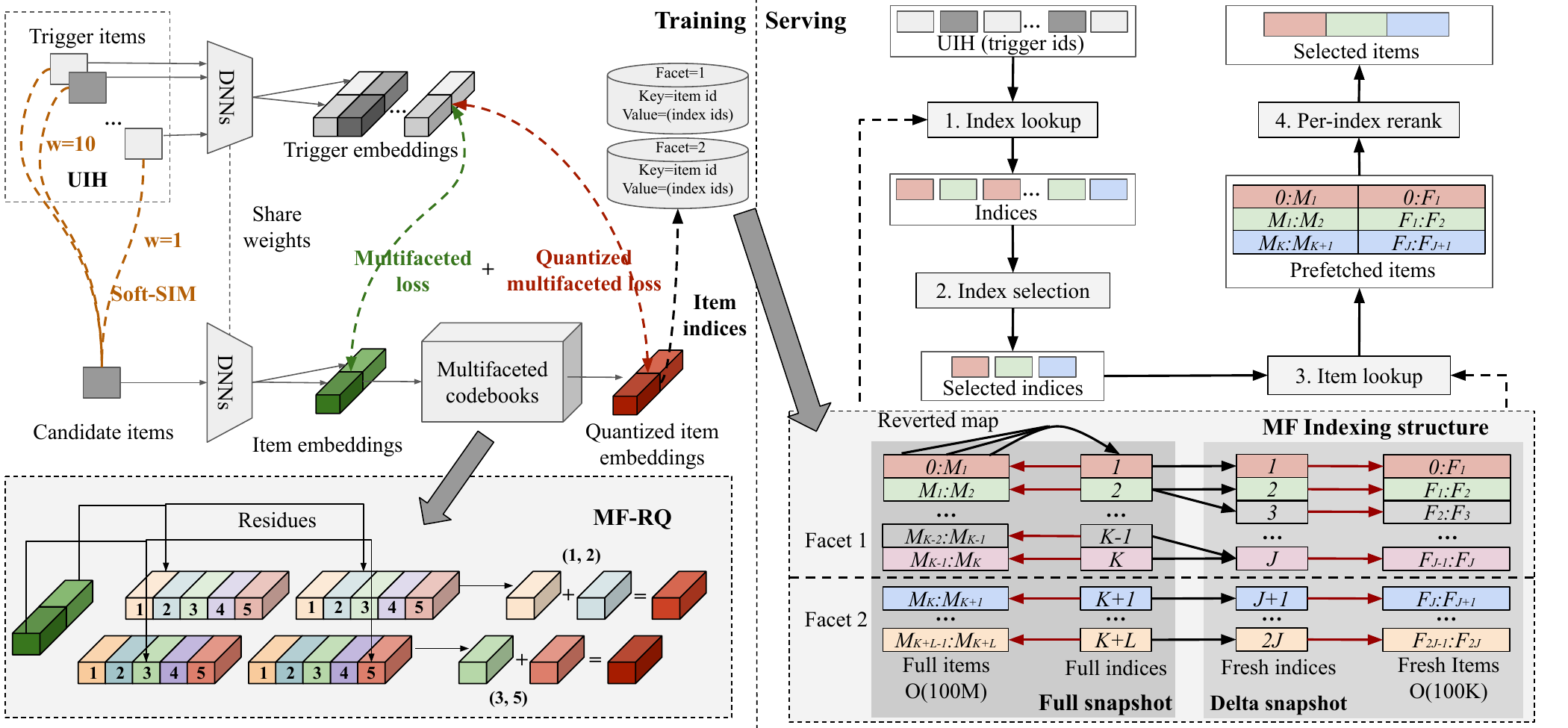}
    \vspace{-2mm}
    \caption{Overview of \ToolX~ framework. During training phase, it learns a multifaceted embedding for each item, and conducts residual quantization on candidate item embeddings by facet in parallel. The codebook and item embeddings are jointly trained via a multifacted loss and a quantized multifaceted loss. The serving part consists of four modules: 1) index lookup, which looks up the multifaceted indices of query item ids; 2) index selection module that selects top $K$ indices; 3) item lookup, which fetches all items in each selected index; 4) per-index rerank, which selects top $N$ items for each selected index. }
    \label{fig:main}
    \vspace{-3mm}
\end{figure*}

\noindent \textbf{Multifaceted Embedding.} Prior works  have shown that a single embedding cannot capture different facets of item correlations. Therefore, multifaceted embeddings can help improve recall during retrieval \cite{zhang2025optimizing,risch2021multifaceted, zhang2022deep,han2022multi}. However, one practical challenge of using multifaceted embeddings for large-scale retrieval is the serving cost, which grows linearly with the number of facets and can become unaffordable in production. This paper addresses this challenge by jointly learning multifaceted item indices and embeddings online, such that the serving overhead is minimal.

\noindent \textbf{Semantic IDs.} Recent works on generative retrieval have presented different approaches based on vector quantization \cite{van2017neural,lee2022autoregressive,luo2025qarm} for building semantic IDs for items, such that the model can be trained to predict semantic IDs with better generalization capability \cite{rajput2023recommender,ju2025generative,deng2025onerec,zhou2025onerec}. There are several key differences between semantic IDs and item indices.
First, semantic IDs typically require no hash collisions \cite{singh2024better}, meaning each item should have a unique semantic ID. By contrast, a learnable index is designed to assign a moderate number of co-engaged items to the same indices to enable effective retrieval.
More importantly, the semantic ID of each item is static during model training \cite{xu2025mmq}. However, the indices of items in \ToolX~ will change fast as real-time user engagement behavior shifts, capturing real-time engagement patterns. This poses significantly more challenges for online indexing learning, such as how to balance the index distribution online and how to guarantee cluster sizes for real-time retrieval serving.
\vspace{-3mm}
\section{Methodology}
\subsection{Overview}

The overall training and serving workflow of \ToolX~ are illustrated in Figure~\ref{fig:main}. Specifically, \ToolX~ designs a multifaceted item representation to capture multiple aspects of the item (e.g., engagement, relevance, freshness), which is trained with a multifaceted loss between trigger-item embeddings and candidate-item embeddings, and a quantized multifaceted loss between trigger-item embeddings and quantized candidate-item embeddings (the codebook). All parameters, including the codebook, are differentiable and optimized via stochastic gradient descent~\cite{ruder2016overview}.
At serving time, \ToolX~ runs a four-step pipeline: 1) index lookup, which identifies the multifaceted indices for each query/trigger item; 2) index selection, which chooses the indices that best capture user interest; 3) item lookup, which retrieves items associated with the selected indices; and 4) per-index reranking, which selects the top items from each index.

To make \ToolX~ robust and deployable at scale, we address several key challenges: 1) learning a stable multifaceted codebook efficiently under online learning (Section~\ref{subsec:train}); 2) bounding the learned cluster sizes to avoid overly large or overly small indices at serving time (Section~\ref{subsec:publish}); 3) designing an indexing structure that supports efficient and scalable multifaceted index lookup and item retrieval (Section~\ref{subsec:index}); and 4) updating the indexing structure in real time with consistency guarantees (Section~\ref{subsec:update}).

\subsection{Training}
\label{subsec:train}
This section presents how the multifaceted hierarchical codebook and item embeddings are jointly learned. We first introduce \emph{multifaceted residual quantization} (MF-RQ) to quantize each facet independently with an $L$-layer hierarchical codebook, and then present the training objective used to optimize both embeddings and codebook parameters end-to-end. Finally, we summarize the techniques we use to improve index balance and stabilize online training.

\noindent \textbf{MF-RQ.}
Differing from traditional RQ \cite{lee2022autoregressive} that uses a two dimensional hierarchical codebook to quantize item embeddings, MF-RQ employs a three-dimensional hierarchical codebook. 
Formally, a $F$-faceted and $L$-layer codebook can be denoted by $\{C_1,\ldots,C_L\}$, where $C_l \in \mathbb{R}^{F \times N_l \times d}$ is the $l$th-layer codebook with size $N_l$ and dimension $d$. Given an item embedding $v \in \mathbb{R}^{F \times d}$, MF-RQ performs residual quantization on the $f$-th facet of an item embedding ($v[f]$) using the corresponding facet-specific codebook ($\{C_1[f],\ldots,C_L[f]\}$). In this way, different facets of item embeddings can be quantized independently and in parallel (see bottom left of Figure~\ref{fig:main}). 
Specifically, we initialize the residual as $r_0 = v$ and, for each layer $l \in \{1,\ldots,L\}$ and facet $f \in \{1,\ldots,F\}$, we select the nearest codeword from the $f$-th sub-codebook at layer $l$ as:
\begin{equation}
\small
\label{eq:mfrq-assign}
k_{l,f} = \arg\min_{k \in \{1,\ldots,N_l\}} \left\| r_{l-1}[f] - C_l[f,k] \right\|_2^2,
\end{equation}
where $r_{l-1}[f] \in \mathbb{R}^d$ denotes the $f$-th facet residual before layer $l$, and $C_l[f,k] \in \mathbb{R}^d$ is the $k$-th codeword for facet $f$ at layer $l$. The layer-$l$ quantized embedding and the updated residual are then given by:
\begin{equation}
\small
\label{eq:mfrq-residual}
q_l[f] = C_l[f,k_{l,f}], \qquad r_l[f] = r_{l-1}[f] - q_l[f].
\end{equation}

After $l$ layers, the quantized embedding is obtained by summing the selected codewords across all previous layers:
\begin{equation}
\small
\label{eq:mfrq-recon}
\hat{v}^l[f] = \sum_{k=1}^{l} q_k[f], \qquad \hat{v}^l \in \mathbb{R}^{F \times d}.
\end{equation}
This computation can be efficiently parallelized on GPU, allowing the indices for all facets to be computed simultaneously.


\noindent \textbf{Training Loss.} To jointly learn item embeddings and indices, we minimize a multifaceted loss between trigger item embeddings and candidate item embeddings, and the same multifaceted loss between trigger item embeddings and quantized candidate item embeddings. Specifically, trigger items are previously engaged items selected from the user interaction history (UIH), while candidate items are the most recently engaged items. We use the Search-based Interest Model (SIM \cite{pi2020search} to identify semantically relevant items within the user’s UIH and assign them higher training weights (see Figure \ref{fig:main}). Moreover, we leverage Sampled SoftMax (SSM) loss \cite{wu2024effectiveness} as the retrieval loss train \ToolX. Given a co-engaged item embedding pair $(v_i, v_j)$ and a set of randomly negative item embedding set $V_k$, the SSM loss for $(v_i, v_j)$ is defined as:
\begin{equation}
\begin{split}
\small
\label{eq:engage-loss}
    L^{s}(v_i,v_j,V_k) \!=\! -\!\sum_{f=1}^{F}\!w_{i,j}[f]\log\frac{e^{v_{i}^T[f]v_{j}[f]}}{e^{v_{i}^T[f] v_{j}[f]} + \sum_{v\in V_k}e^{v_{i}^T[f]v[f]}},
\end{split}
\end{equation}
where $v_i[f]\in\mathrm{R}^d$, $v_j[f]\in\mathrm{R}^d$, $v[f]\in\mathrm{R}^d$ represents the $f$-th embedding of $v_i$, $v_j$, $v$ respectively, $d$ is item embedding dimension, $F$ is the total number of facets, and $w_{i,j}[f]$ is the $f$-th co-engagement label between item $i$ and $j$. Note that different facets of item embeddings and codebook are trained on different labels. In addition, we compute the SSM loss between $(v_i,\hat{v}_j^{\,l})$ as $L^{\mathrm{ssm}}(v_i,\hat{v}_j^{\,l},V_k)$, where $\hat{v}_j^{\,l}$ denotes the quantized embedding of item $j$ after the $l$-th quantization layer. The overall training objective for the pair $(v_i, v_j)$ is defined as:
\begin{equation}
\small
\label{eq:engage-loss}
L^{e}(v_i,v_j,V_k)
\!=\!
w_0\,L^{\mathrm{s}}(v_i,v_j,V_k)
\!+\!
\sum_{l=1}^{L}\!w_l\,L^{\mathrm{s}}(v_i,\hat{v}_j^{\,l},V_k)
\!+\!
L^{\mathrm{a}}(v_i,v_j,V_k),
\end{equation}
where $w_0,w_1,\ldots,w_L$ are tunable weights and $L^{\mathrm{a}}$ denotes auxiliary losses from other non-engagement tasks. In this paper, we incorporate a semantic relevance loss into $L^{\mathrm{aux}}$ and apply it to the second facet to improve semantic relevance, referring to the design in \cite{zhang2025optimizing}. 

\noindent \textbf{Index Balance Control.} We employ two main index-balancing strategies. First, we adopt a \textit{delayed start} scheme by warming up item embeddings before activating the codebook and commencing joint training. In addition, we progressively activate the codebook in a layer-wise manner to improve the stability and robustness of multi-layer codebook training. Second, we introduce a \textit{regularization term} that penalizes over-utilized (i.e., overly popular) codewords to encourage balanced codeword usage (see Appendix \ref{apx:param} for details). Note that for codebook initialization, we randomly sample a set of item embeddings and use them to initialize the codebook layer by layer. 

\vspace{-1mm}
\subsection{Index Rebalance}
\label{subsec:publish}
\begin{figure}[t]
    \centering
    \includegraphics[width=0.98\linewidth]{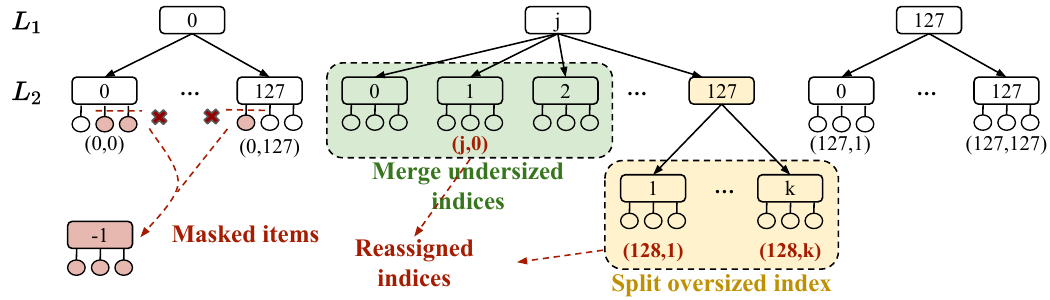}
    \vspace{-0.1in}
    \caption{Illustration of index rebalancing, where undersized indices are merged, oversized indices are split, and mask items are assigned with an invalid index.}
    \label{fig:split}
    \vspace{-3mm}
\end{figure}

While we develop several techniques to improve codebook balance during training, these methods do not directly guarantee bounded index sizes at serving time. As a result, some indices can still become overly large or overly small in production, which degrades retrieval quality and efficiency. This issue is further exacerbated in online recommendation systems, where distribution shift between the training item set (primarily exposed items) and the serving-time candidate pool (exposed + unexposed items) makes it difficult to control the serving-time index distribution \cite{yang2023generic,zhang2024robust}.
To address this, we introduce a index-rebalance step, \emph{split-and-merge}, which splits oversized indices and merges undersized ones as shown in Figure~\ref{fig:split}. This procedure guarantees that the number of items assigned to each index remains bounded in production. In addition, we design a flexible per-facet index-pruning mechanism for \ToolX, enabling us to easily customize the candidate pool for different facets.

\noindent\textbf{Split-and-Merge with Index-Size Guarantees.}
For each oversized index (size $> B_{\mathrm{upp}}$), we independently run $k$-means clustering on the last-layer residual item embeddings associated with that index, so the splitting step can be fully parallelized. Suppose splitting produces $N_s$ new centroids. We then increase the total number of original indices from $N$ to $N + N_s$. If an item is assigned to new centroid $i$, we reset its index to $N + i$. Moreover, to prune undersized indices, we scan all $L$-th-layer indices under each $(L-1)$-th-layer index and identify those with size below $B_{\mathrm{low}}$. We then merge these small indices by reassigning their items to the smallest index among them. In the rare case where the merged index remains smaller than $B_{\mathrm{low}}$, we further merge it with a nearby index that is not oversized.

\noindent\textbf{Item Pruning for Candidate Set Customization.}
With \ToolX, we can flexibly derive a subset of the full candidate corpus for each index facet, enabling facet-specific customization of retrieved items. For example, suppose the second facet is trained with an auxiliary semantic-relevance loss to group semantically related items into the same index, which is beneficial for cold-start items. We can then customize the pool by assigning cold-start items only to this second facet. Specifically, as illustrated in bottom left of Figure \ref{fig:split}, we extend the total number of indices by one to represent an \emph{invalid} index, and reassign all masked items to this invalid index, so they are excluded from retrieval at serving time.

\vspace{-1mm}
\subsection{Indexing Structure}
The indexing structure in \ToolX~ primarily consists of two mappings: an item-to-index mapping and an index-to-item mapping. These mappings are stored as in-memory tensors to enable efficient lookups. We describe the details below.

\label{subsec:index}
\noindent \textbf{Unified Index.} In \ToolX, an item is assigned with $F$ indices after RQ, one per facet, each of which is originally represented as a tuple $(c^{f}_{1},\ldots,c^{f}_{L})$. Therefore, we first cover them into unified integer indices as:
\begin{equation}
\small
    c^f = \sum_{1}^{L-1}c_l^f*(\Pi_{k=1}^{L-l}N_k) + c_L^f,
\end{equation}
Suppose facet $f$ has $M_f$ distinct index values in total after index-rebalancing. We then define a \emph{unified index} vector by offsetting the index of each facet into a single contiguous range:
\begin{equation}
\small
\label{eq:unified-index}
\tilde{c}(v)
=
\bigl[\,
c^{1},\;
M_{1}+c^{2},\;
\ldots,\;
\textstyle\sum_{f=1}^{F-1} M_{f} + c^{F}
\,\bigr]
\in \mathbb{R}^{F},
\end{equation}
where $c^{f}$ denotes the (scalar) merged index for facet $f$, and $M_f$ is the total number of merged indices for facet $f$.

\noindent \textbf{Item-to-index Mapping.} With unified index, we maintain an item-to-index mapping using (i) a dense tensor that stores the unified indices for all items with size $(I,F)$, and (ii) a lightweight \emph{direct lookup} structure that maps each item id to its row offset $i\in[0,I)$, where $I$ is the pool size. Given an item id, we first retrieve its row offset in constant time, and then fetch the corresponding multifaceted indices via a single indexed read from the $(I,F)$ tensor. The time and space complexity are $O(1)$ and $O(I\times F)$ respectively. 

\noindent \textbf{Index-to-item Mapping.}
\ToolX~ stores the index-to-item mapping using two tensors: (i) a count tensor $T^{\mathrm{index}} \in \mathbb{N}_{+}^{\sum_{f=1}^{F} M_f}$, where $T^{\mathrm{index}}[m]$ is the number of items assigned to the $m$-th unified index and $\sum_{f=1}^{F} M_f$ is the total amount of indices; and (ii) an item-id tensor $T^{\mathrm{item}} \in \mathbb{N}^{I}$, which stores all item ids sorted by unified indices. Consequently, the items under the $m$-th unified index correspond to the tensor segment
\begin{equation}
\small
\label{eq:index-to-item-range}
T^{\mathrm{item}}
\Bigl[
\textstyle\sum_{k=1}^{m-1} T^{\mathrm{index}}[k]
\;:\;
\sum_{k=1}^{m} T^{\mathrm{index}}[k]
\Bigr].
\end{equation}
The time and space complexity for finding items under an index are $O(1)$ and $O(I + \sum_{f=1}^{F} M_f)$ respectively. Note that details on how we use these mappings in serving are shown in Section \ref{subsec:serving}.

\begin{figure}
    \centering
    \includegraphics[width=0.92\linewidth]{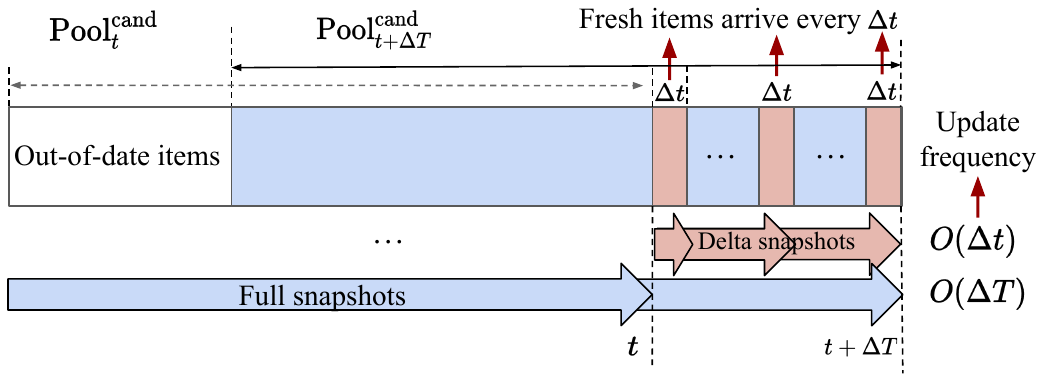}
    \vspace{-3mm}
    \caption{Delta update of \ToolX’s indexing structure. Every $\Delta t$ time (e.g., 1 min), a delta snapshot storing the indexing structure of fresh items is produced. Every $\Delta T$ time (e.g., 30 minutes), a full snapshot storing the refreshed indexing structure of the full item pool is produced. \ToolX~ fetches items from both the full and delta snapshots during serving.}
    \label{fig:update}
    \vspace{-4mm}
\end{figure}
\vspace{-1mm}
\subsection{Real-time Update}
\label{subsec:update}
In online training, both item embeddings and item inventory evolve continuously, hence it is important to keep \ToolX's indexing structure up to date in real time. To this end, we design a \emph{delta-update} strategy: a \emph{delta snapshot} that indexes newly arrived items can be produced within $O(\mathrm{min})$, while a \emph{full snapshot} that indexes the entire item pool is refreshed periodically. At serving time, \ToolX\ fetches items under the selected indices from both the full and delta snapshots and then merges the retrieved candidates (Figure \ref{fig:update}).

\noindent\textbf{Motivation.} Conceptually, there are two approaches to updating the indexing structure. The first is an \emph{asynchronous in-training update}: we maintain a cached mapping from items to indices and update it whenever an item is re-embedded and re-quantized during training. This approach guarantees the freshness of indexing structure, but it has two major drawbacks. First, because the updates are asynchronous with respect to the checkpoint used for serving, the published indices can become inconsistent across items at serving time. Second, training data typically covers only impression (exposed) items; consequently, unexposed or newly arrived items in the item pool may never be refreshed through training-time updates. Alternatively, the second approach is to periodically recompute indices for the item pool using a fixed model checkpoint. This yields mutually consistent embeddings, codebook, and indices, but incurs higher compute cost and longer end-to-end refresh latency, delaying the delivery of fresh items.

\noindent\textbf{Design Details.} In practice, we adopt the second approach and mitigate its freshness limitation with our delta-update design: periodic full-snapshot refreshes ensure global consistency of the indexing structure, while real-time delta snapshots ensure newly arrived items become retrievable within $O(\mathrm{min})$. Since the number of fresh items in the delta snapshot is much smaller than the size of the full candidate set (roughly 100$\times$), \ToolX~ skips index-rebalancing for fresh indices and items, and the computational cost of producing the delta snapshot is minimal. 
Note that the fresh indices are the original mapped codeword indices of fresh items. However, this can introduce some inconsistencies between the indexing structures of the full and delta snapshots, since the full snapshot may end up with more indices than the delta snapshot due to index-rebalancing. To address this, we maintain a mapping from fine-tuned indices to their original indices, as shown in bottom right of Figure \ref{fig:main}. At serving time, once an index from the full snapshot is selected, we directly look up all associated fresh indices and fetch all items contained in those indices (more details are provided in Section \ref{subsec:serving}).
\begin{table*}[t]
\centering
\small
\caption{Performance comparison of \ToolX\ and baseline methods. For recall, we report recall@1000 on VVC (Video View Complete), Like, LWT (Long Watch Time), and CCD (Cold Content Delivery) tasks. For semantic relevance, we report the average topic match rate between the trigger and retrieved items at the $T_1$ (coarse-grained) and $T_2$ (fine-grained) levels. }
\vspace{-0.1in}
\label{tab:main}
\begin{tabular}{p{0.7in}|p{0.85in}p{0.85in}p{0.85in}p{0.85in}|p{0.85in}p{0.85in}}
\toprule
\multirow{2}{*}{Baseline} & \multicolumn{4}{c}{Recall} & \multicolumn{2}{c}{Semantic Relevance} \\ \cline{2-7}
 & VVC & Like & LWT & CCD & T1 topic & T2 topic \\
\midrule
NeuCF~\cite{he2017neural}          & 21.51\%        & 24.44\%
                                   & 25.85\%        & 7.54\%
                                   & 24.80\%        & 17.10\% \\

MoL~\cite{zhai2023revisiting}      & 22.67\% (+5.39\%)        & 25.78\% (+5.48\%)
                                   & 26.84\% (+3.83\%)        & 7.60\%  (+0.80\%)
                                   & 39.65\% (+59.88\%)       & 26.84\% (+56.96\%) \\

HLLM~\cite{chen2024hllm}           & 22.39\% (+4.09\%)        & 25.28\% (+3.44\%)
                                   & 26.61\% (+2.94\%)        & 7.97\%  (+5.70\%)
                                   & 38.94\% (+57.02\%)       & 26.48\% (+54.85\%) \\
HSTU~\cite{zhai2024actions}          & \underline{23.64\% (+9.90\%)}        & \textbf{28.07\% (+14.85\%)}
                                   & 26.46\% (+2.36\%)        & \textbf{12.08\% (+60.21\%)}
                                   & 28.00\% (+12.90\%)       & 19.07\% (+11.52\%) \\
MTMH~\cite{zhang2025optimizing}    & 22.85\% (+6.23\%)        & 26.11\% (+6.83\%)
                                   & \underline{27.23\% (+5.34\%)}        & 8.18\%  (+8.49\%)
                                   & 46.10\% (+85.89\%)       & 33.55\% (+96.20\%) \\
VQIndex~\cite{bin2025real}         & 19.41\% (-9.76\%)        & 22.77\% (-6.83\%)
                                   & 22.98\% (-11.10\%)       & 7.60\%  (+0.80\%)
                                   & \underline{47.47\% (+91.41\%)}       & \underline{34.56\% (+102.11\%)} \\

\textbf{MFLI (Ours)}               & \textbf{24.04\% (+11.76\%)} & \underline{26.72\% (+9.33\%)}
                                   & \textbf{27.49\% (+6.34\%)}  & \underline{11.86\% (+57.29\%)}
                                   & \textbf{52.31\% (+110.93\%)}& \textbf{38.70\% (+126.32\%)} \\
\bottomrule
\end{tabular}
\vspace{-4mm}
\end{table*}

\vspace{-5mm}
\subsection{Serving}
\label{subsec:serving}
The overall serving flow of \ToolX{} follows an item-to-item retrieval paradigm, as demonstrated in Figure \ref{fig:main}. Upon receiving a user request, we extract a set of $T$ engaged items from the user's interaction history as triggers. 
Each trigger item is first mapped to a multifaceted index vector $\tilde{c}$ via the \textit{index lookup module}. Next, the \textit{index selection module} selects the indices that best capture the user's current interests. The \textit{item selection module} then retrieves items from the selected indices. Finally, a reranker performs \textit{per-index reranking} to select the top-$k$ candidates for each index and merges the results across indices. We describe each module in detail below.

\noindent \textbf{Index Lookup.}
For each trigger item, this module retrieves its mapped unified index vector $\tilde{c}(t)$ using the \textit{item-to-index mapping} (see Section~\ref{subsec:index}).
Concretely, given an item ID, its row offset $i \in [0, I)$ is first obtained via a direct-lookup structure, and then the corresponding $\tilde{c}(t)$ is fetched with a single indexed read from the cached item-to-index mapping tensor. The total time complexity for each request is $O(T)$, and $T*F$ indices will be found.

\noindent \textbf{Index Selection.} After receiving $T*F$ indices from the index lookup module, the index selection module selects $K$ unique indices, each of which contains a set of correlated items that a user may be interested in.
Because multiple trigger items can map to the same single-facet index, we construct an index-frequency histogram per facet, which can be used for multi-user-interest mining \cite{yan2024trinity}.
We describe several strategies used in \ToolX~ (additional detail in Appendix~\ref{apx:index}):
\vspace{-1mm}
\begin{itemize}[leftmargin=*]
    \item \textit{Multi-interest extraction}: \ToolX{} applies multinomial sampling \cite{glick1973sample} over the index-frequency distribution, so indices mapped from more trigger items are more likely to be selected. Unless otherwise specified, we use this strategy to sample top-$k$ indices in subsequent experiments.
    \item \textit{Recent-interest boosting}: for indices associated with the most recent triggers, \ToolX~ increases their priority to capture users' latest interests more responsively.
    \item \textit{Long-tail interest exploration}: for long-tail users with limited interaction history, fewer trigger items are extracted, which yields fewer mapped indices. To mitigate this, for each selected index, we additionally include neighboring indices that share the same upper-level indices, enabling more diversity for exploration.
\end{itemize}

\noindent\textbf{Item Selection.}
This module takes as input the $K$ selected unified indices and retrieves all items associated with them from both the \textit{full snapshot} and the \textit{delta snapshot}.
For a given index $i$, the module first fetches the associated items from the latest \textit{full snapshot}, denoted by $\mathcal{C}^{\mathrm{full}}_i$ (see Eq.~\eqref{eq:index-to-item-range}).
Next, it performs a direct lookup to obtain the \emph{fresh} index (or indices) associated with $i$.
Without index-rebalancing, the associated fresh index is simply $i$ itself.
With index-rebalancing, if $i$ is produced by splitting an originally oversized index, then the fresh index corresponds to the original (pre-split) index; if $i$ is produced by merging a set of undersized indices, then the associated fresh indices correspond to those original undersized indices.
After extracting the fresh indices, the module fetches the corresponding fresh items from the latest \textit{delta snapshot}, denoted by $\mathcal{C}^{\mathrm{delta}}_i$.
The final set of items retrieved is $\mathcal{C}=\bigcup_{i \in \{1,\ldots,K\}} \mathcal{C}_i$,
where $\mathcal{C}_i=\mathcal{C}^{\mathrm{full}}_i \cup \mathcal{C}^{\mathrm{delta}}_i$ denotes the retrieved items from index $i$, and $|\mathcal{C}|=O(K*B_{upper})$ is bounded due to split-and-merge step.

\noindent\textbf{Per-index Reranking.}
\label{subsubsec:serving:per-index-rerank}
This module performs \emph{per-index} reranking using a user-to-item reranker model \cite{zhai2023revisiting}.
For each selected index, it ranks the items retrieved in the previous step (in parallel across indices) and retains the top-$N$ items for that index.
The resulting per-index shortlists are then merged into a single set, which is passed to subsequent ranking stages. With per-index reranking, \ToolX~ preserves multi-interest coverage across indices while ensuring strong trigger-to-item relevance in retrieval. This design helps prevent any single dominant interest from overwhelming the candidates, improving their quality and diversity\cite{li2024full}.
\begin{table*}[t]
\centering
\small
\caption{Ablation study results of \ToolX. We independently remove split-and-merge index rebalancing, top-k index selection, and delayed start to quantify the impact of each component on \ToolX’s recall and semantic relevance. VVC, Like, LWT, and CCD denote the video view complete, like, long watch time, and cold content delivery tasks, respectively.}
\vspace{-0.1in}
\label{tab:ablation}
\begin{tabular}{p{0.8in}|p{0.8in}p{0.8in}p{0.8in}p{0.85in}|p{0.9in}p{0.9in}}
\toprule
\multirow{2}{*}{Baseline} & \multicolumn{4}{c}{Recall} & \multicolumn{2}{c}{Semantic Relevance}\\ \cline{2-7}
 & VVC & Like & LWT & CCD & T1 topic & T2 topic \\
\midrule
\underline{\ToolX}                 & \underline{24.04\%}        & \underline{26.72\%}
                                  & \underline{27.49\%}        & \underline{11.86\%}
                                  & \underline{52.31\%}        & \underline{38.70\%} \\
- Split-and-merge      & 19.60\% (-18.47\%)        & 23.05\% (-13.74\%)
                       & 22.65\% (-17.61\%)        & 11.32\% (-4.55\%)
                       & 49.35\% (-5.66\%)         & 34.49\% (-10.88\%) \\

- Index selection      & 21.61\% (-10.11\%)        & 25.15\% (-5.88\%)
                       & 25.30\% (-7.97\%)         & 12.50\% (+5.40\%)
                       & 53.39\% (+2.06\%)         & 40.17\% (+3.80\%) \\

- Delayed start          & 23.60\% (-1.83\%)         & 26.38\% (-1.27\%)
                       & 27.64\% (+0.55\%)         & 12.24\% (+3.20\%)
                       & 51.70\% (-1.17\%)         & 38.28\% (-1.09\%) \\

\bottomrule
\end{tabular}
\vspace{-2mm}
\end{table*}

\vspace{-1mm}
\section{Experiments}
\label{sec:eval}
In this section, we evaluate \ToolX~ on a commercial recommendation system. We first compare \ToolX's offline performance with prior state-of-the-art industry retrieval models. Next, we present an ablation study of \ToolX~ and provide a comprehensive analysis of the learned index distribution. We then assess the impact of key hyperparameters on performance. Finally, we deploy \ToolX~ and benchmark it against the baseline retrieval model in online experiments. More training data details are presented in Appendix~\ref{apx:param}.

\vspace{-1mm}
\subsection{Baselines and Evaluation Metrics}
\label{subsec:baseline}
For offline evaluation, we compare \ToolX~ against six baseline models in terms of recall and item semantic relevance. Specifically, we report recall@1000 on four tasks: Video View Complete (VVC), Like, Long Watch Time (LWT), and Cold Content Delivery (CCD). We define semantic relevance as the average topic match rate between the input trigger items and the retrieved candidate items. We consider two types of topic IDs, denoted by $T_1$ and $T_2$: $T_1$ captures broader item topics, whereas $T_2$ captures more fine-grained topics. Unless otherwise specified, \ToolX~ uses two facets. Each facet employs a two-layer codebook with sizes 512 and 128, respectively. The first facet is trained using only a co-engagement loss, while the second facet is trained with both a co-engagement loss and a relevance loss \cite{zhang2025optimizing}. Moreover, we train \ToolX\ and baselines using real data from Period~$P_1$, and evaluate all models on real data from Period~$P_2$, which follows $P_1$ (see Appendix \ref{apx:param} for additional details on model hyperparameters and training setups).
We describe the compared baselines in detail below:
\begin{itemize}[leftmargin=*]
    \item \textbf{NeuCF}~\cite{he2017neural}: it uses only content-agnostic ID features as input and produces a single-facet item embedding via a sparse neural network (NN) followed by a dense NN.
    \item \textbf{MoL}~\cite{zhai2023revisiting}: it extends NeuCF by additionally incorporating item content features as inputs to the DNN.
    \item \textbf{HLLM}~\cite{chen2024hllm}: it augments the item tower with content embeddings generated by pretrained LLMs.
    \item \textbf{VQIndex}~\cite{bin2025real}: it adapts streaming VQ on item retrieval and learns a single-layer codebook of size 10k during training.
    \item \textbf{HSTU}~\cite{zhai2024actions}: It is a state-of-the-art sequential recommendation model based on Transformers. For evaluation, we use its learned item embeddings to perform item-to-item retrieval.
    \item \textbf{MTMH}~\cite{zhang2025optimizing}: It is the state-of-the-art retrieval model based on a multi-head architecture with multi-task learning.
\end{itemize}
Note that VQIndex is also an index-based retrieval approach with single facet, while all other baselines are ANN-based methods.

\vspace{-1mm}
\subsection{Main Results}
\label{subsec:main}
We first compare the offline evaluation results of \ToolX~ against baselines on four engagement tasks: VVC, LWT, and CCD, as well as two item semantic relevance tasks: $T_1$ and $T_2$ topic match rates. 
As demonstrated in Table \ref{tab:main}, \ToolX~ achieves both top recalls and relevance across various tasks.
Specifically, among the four engagement tasks, \ToolX~ has the highest recall on VVC and LWT, with up to +11.76\% and +6.34\% improvements respectively. For Like and CCD, \ToolX~ achieves 26.72\% and 11.86\%, outperforming all other baselines, expect for HSTU. However, HSTU has significantly worse semantic relevance rate. This is expected since the item embeddings of HSTU is trained to maximize the retrieval recall on co-engagement tasks, without explicitly optimizing the item semantic relevance tasks.

Moreover, \ToolX{} outperforms all baselines on semantic relevance, with up to +110.93\%/+126.32\% gains in $I_1$/$I_2$ topic match rate. This shows \ToolX{} retrieves more co-engaged items while preserving intra-index semantic coherence, improving interest matching. Finally, our \ToolX~ delivers higher end-to-end throughput than ANN-based retrieval models. Particularly, compared with MTMH, a multi-ANN retrieval model, \ToolX{} increases throughput by 60\%.
%

\subsection{Ablation Study}
\label{subsec:ablation}
Next, we conduct an ablation study on \ToolX~ to quantify the impact of three design choices: split-and-merge index-rebalancing, top-$k$ index selection, and delayed start. We report offline recall on four engagement tasks (VVC, Like, LWT, CCD) and item semantic relevance measured by $T_1$/$T_2$ topic match rates.

As shown in Table~\ref{tab:ablation}, \textit{split-and-merge} significantly improves \ToolX's performance. Removing it leads to consistent regressions in both recall and semantic relevance (e.g., $-18.47\%$ Recall on VVC and $-10.88\%$ $T_1$ match rate). This is expected: pruning overly small indices avoids retrieving too few items, while splitting overly large indices improves intra-index homogeneity so that each index corresponds to a tighter cluster of similar items. Moreover, removing top-$k$ index selection improves semantic relevance ($T_1$/$T_2$: $+2.06\%$/$+3.80\%$) but reduces recall on the primary engagement tasks (e.g., $-10.11\%$ and $-5.88\%$ on VVC and Like), indicating a trade-off between engagement-oriented retrieval effectiveness and semantic alignment. 
Finally, removing delayed start reduces recall on VVC and Like, while yielding only marginal gains in semantic relevance.

\begin{figure}[t]
    \centering
    \begin{subfigure}{0.45\linewidth}
        \includegraphics[width=\linewidth]{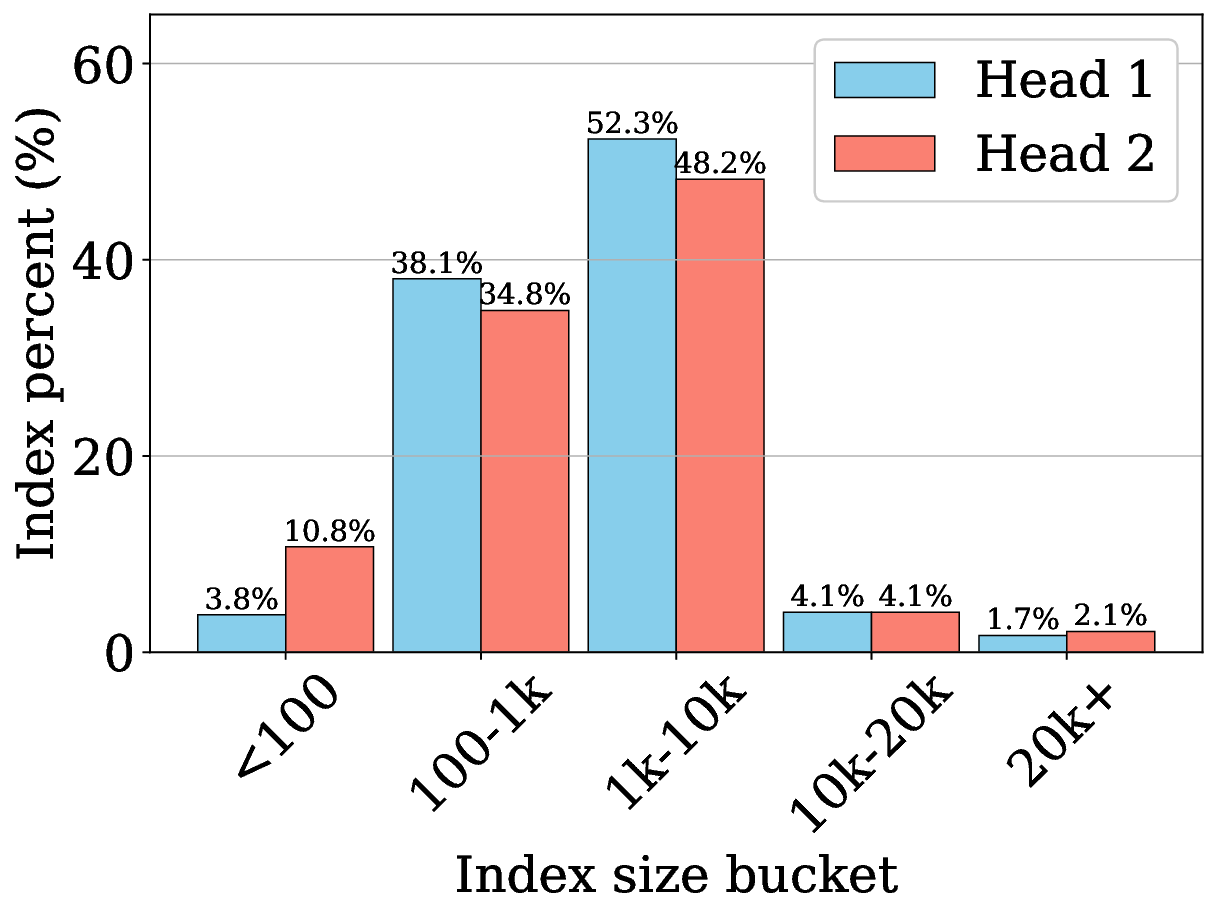}
        \vspace{-5mm}
        \caption{Index size distribution}
        \label{fig:index1}
    \end{subfigure}
    \begin{subfigure}{0.45\linewidth}
        \includegraphics[width=\linewidth]{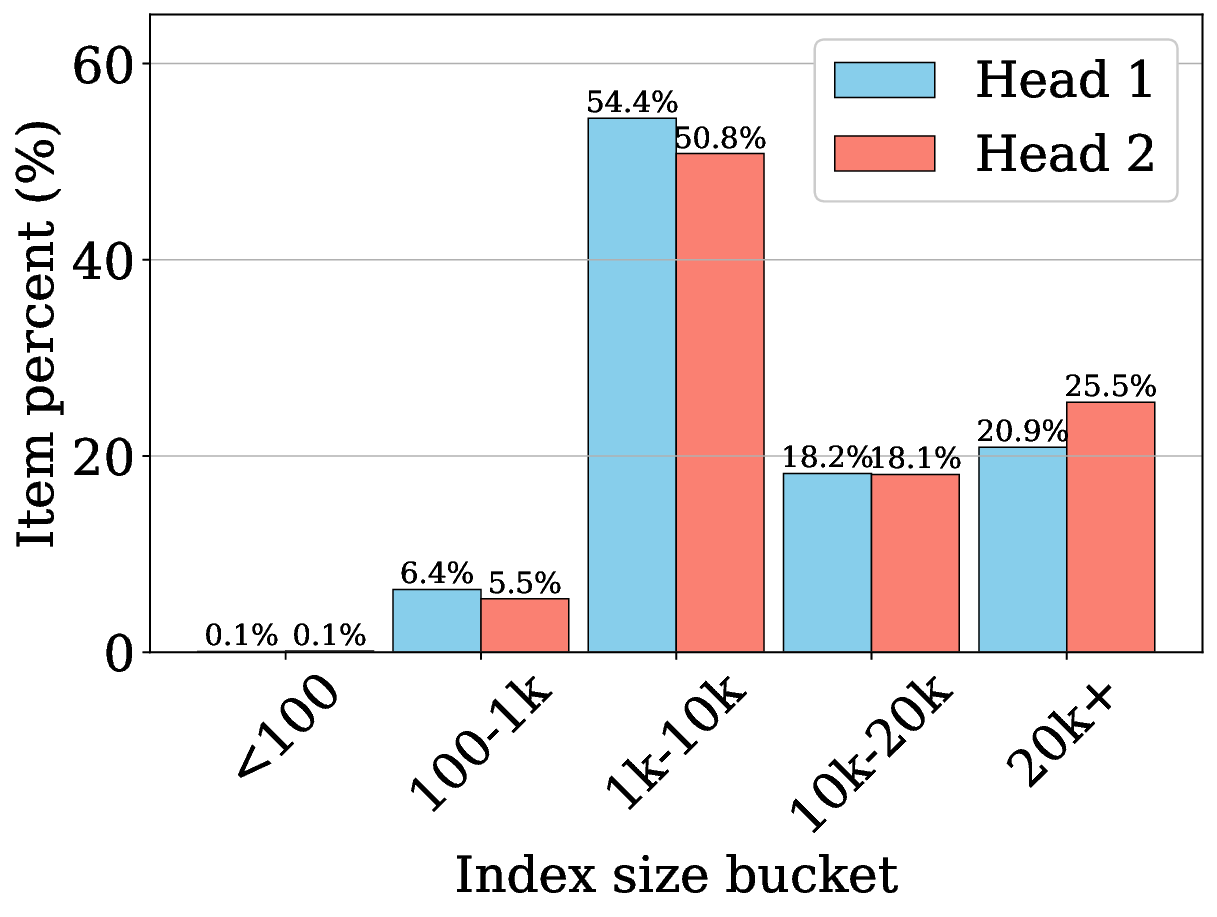}
        \vspace{-5mm}
        \caption{Item size distribution}
        \label{fig:index2}
    \end{subfigure}
    \begin{subfigure}{0.45\linewidth}
        \includegraphics[width=\linewidth]{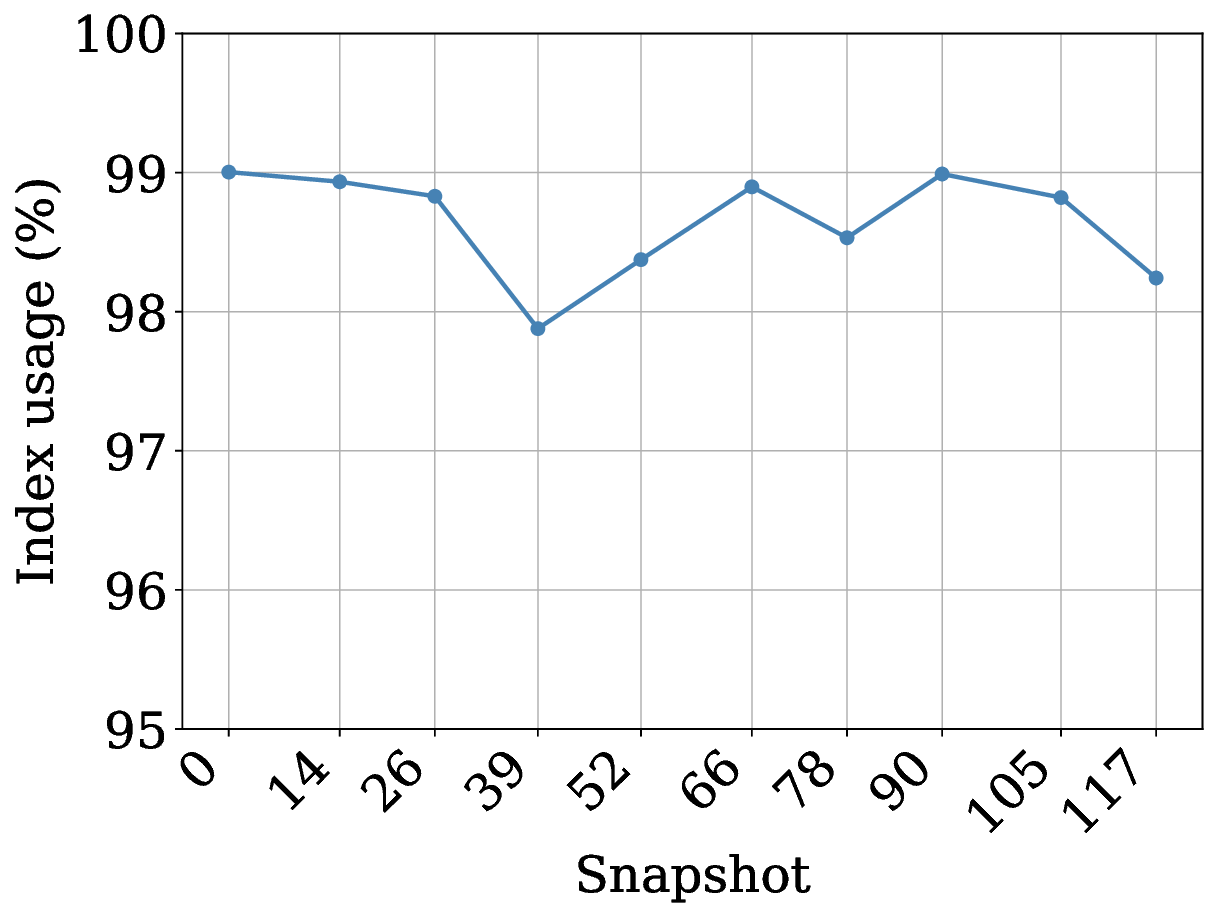}
        \vspace{-5mm}
        \caption{Index usage across snapshot}
        \label{fig:index3}
    \end{subfigure}
    \begin{subfigure}{0.45\linewidth}
        \includegraphics[width=\linewidth]{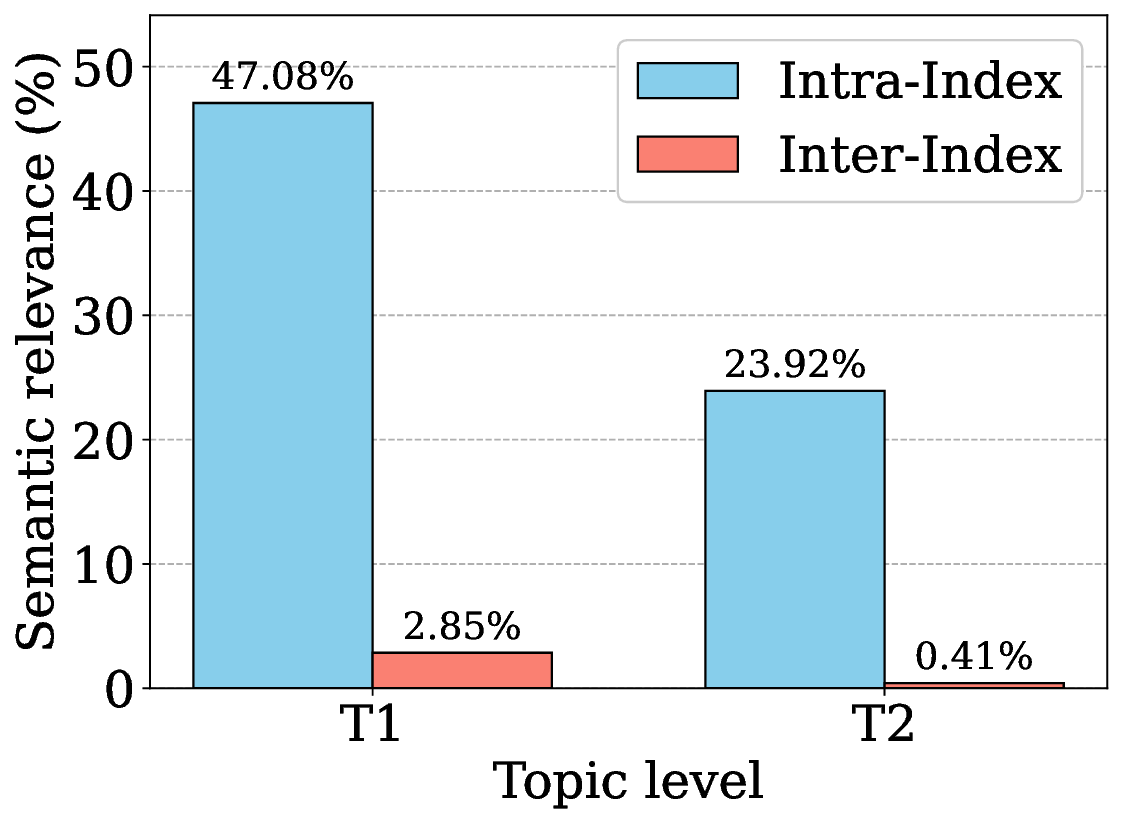}
        \vspace{-5mm}
        \caption{Index-level relevance}
        \label{fig:index4}
    \end{subfigure}
    \vspace{-3mm}
    \caption{\textbf{Analysis of learnt indices in \ToolX.}
    (a) Index-size distribution (items per index).
    (b) Item distribution across index-size buckets (items per index-size buckets).
    (c) Index usage (percentage of non-empty indices) over training snapshots.
    (d) Intra-index vs.\ inter-index relevance.}
    \label{fig:tradeoff}
    \vspace{-5mm}
\end{figure}

\subsection{Index Distribution Analysis}
\label{subsec:analysis}
In this subsection, we conduct a comprehensive evaluation of the quality of the learned indices in \ToolX. We use a two-layer codebook of size $512 \times 128$, which maps approximately 200M items into discrete indices. We first examine the distribution of index sizes. As shown in Figure~\ref{fig:index1}, 98\% of indices contain fewer than 20k items, and approximately 90\% fall between 100 and 10k items. In addition, we analyze the item-level distribution across indices (Figure~\ref{fig:index2}). 

We observe that around 80\% of items are assigned to indices with fewer than 20k items, and more than 60\% are assigned to indices smaller than 10k. Note that these statistics are computed \emph{before} split-and-merge; afterward, overly small (size < 100) and overly large indices (size > 20k) are removed. Moreover, we analyze how index usage evolves during online training over a 7-day window. As shown in Figure~\ref{fig:index3}, more than 98\% of indices are actively used, and this ratio remains stable throughout continuous training. This indicates that \ToolX~ maintains broad index coverage over time and can be trained robustly in online recommendation systems.

Last, we compute intra-index and inter-index semantic relevance to assess the semantic purity of the learned indices. Specifically, we randomly sample item pairs from the same index and compute the average topic match rate as the intra-index relevance. We also randomly sample item pairs from different indices and compute the average topic match rate as the inter-index relevance. As demonstrated in Figure~\ref{fig:index4}, intra-index relevance is substantially higher than inter-index relevance, indicating that items within the same index are indeed semantically related.

\begin{table}[t]
\centering
\small
\caption{Effects of number of facets and codebook size.}
\vspace{-1mm}
\label{tab:ablation-combined}
\begin{subtable}[t]{\linewidth}
\centering
\begin{tabular}{p{0.65in}|p{0.72in}p{0.72in}|p{0.5in}}
\toprule
Num. of facets & Recall of VVC & Recall of LWT & Throughput\\
\midrule
One & 20.72\% (-13.81\%) & 24.49\% (-10.91\%) & +0.00\%\\
\underline{Two} & \underline{24.04\%} & \underline{27.49\%} & \underline{-18.8\%}\\
Three & 24.13\% (+0.37\%) & 28.19\% (+2.55\%) & -23.9\%\\
\bottomrule
\end{tabular}
\caption{Varying number of facets.}
\label{tab:ablation-facet}
\end{subtable}
%
\begin{subtable}[t]{\linewidth}
\centering
\begin{tabular}{p{0.65in}|p{0.7in}p{0.7in}|p{0.75in}}
\toprule
Codebook size & Recall of VVC & Recall of LWT & $T_1$ topic\\
\midrule
10k & 22.54\% (-6.24\%) & 26.71\% (-2.84\%) & 55.10\% (+5.33\%)\\
256$\times$128 & 24.07\% (+0.12\%) & 27.53\% (+0.15\%) & 52.21\% (-0.19\%)\\
\underline{512$\times$128} & \underline{24.04\%} & \underline{27.49\%} & \underline{52.31\%}\\
1024$\times$256 & 23.18\% (-3.58\%) & 27.15\% (-1.24\%) & 52.73\% (+0.80\%)\\
\bottomrule
\end{tabular}
\caption{Varying codebook size.}
\label{tab:ablation-codebook}
\end{subtable}
\vspace{-3mm}
\end{table}

\subsection{Varying Hyperparameters}
\label{subsec:codesize}
In this subsection, we study the sensitivity of \ToolX{} to two key hyperparameters: the \textit{number of facets} and the \textit{codebook size}. We report recall on VVC/CCD and item semantic relevance on $T_1$ topics under different settings in Tables~\ref{tab:ablation-facet} and~\ref{tab:ablation-codebook}, respectively (full results are reported in Appendix \ref{apx:full}).

\noindent \textbf{Number of Facets.} We vary the number of facets from 1 to 3. The first facet is trained on an aggregated engagement objective (i.e., the sum of engagement labels), the second emphasizes item semantic relevance via multi-task learning, and the third is specialized for time-spent-related engagement signals. For a fair comparison, we fix the total number of retrieved items by adjusting the number of selected indices per facet; therefore, increasing the number of facets (heads) does \emph{not} increase the downstream ranking overhead, and only incurs small overhead during index lookup and  selection.

As shown in Table~\ref{tab:ablation-facet}, overall recall improves as the number of facets increases. Using a single facet results in substantial regressions across all metrics (VVC: $-13.81\%$; LWT: $-11.21\%$; semantic relevance: $-12.98\%$), indicating that a single representation is insufficient to capture diverse engagement patterns. Increasing from two to three facets yields a modest gain in VVC ($+0.37\%$) but a larger improvement in LWT ($+3.12\%$), consistent with the third facet being explicitly optimized for time-spent-related objectives. Importantly, because we keep the total number of retrieved items fixed across settings, increasing the number of facets \textit{does not} linearly decrease end-to-end serving throughput (three facets: -23.9\%).

\noindent \textbf{Codebook Size.}
We consider four variants: a single-layer codebook of size 10k, and three two-layer codebooks with increasing capacity. As shown in Table~\ref{tab:ablation-codebook}, the single-layer 10k setting substantially degrades recall (VVC: -6.24\%, LWT: -2.84\%) despite achieving the highest topic relevance (+5.33\%), suggesting that coarse quantization forms semantically coherent but less discriminative clusters. In contrast, among two-layer codebooks, increasing capacity improves semantic alignment with only a slight drop in recall: moving from 512$\times$128 to 1024$\times$256 increases topic relevance (+0.80\%) while slightly reducing recall (VVC: -3.58\%; LWT: -1.24\% $\rightarrow$ 27.15\%). Overall, 512$\times$128 offers a robust balance between recall and relevance.

\subsection{Online A/B Testing}
\label{subsec:online}
We deployed \ToolX\ on a real-world recommendation system and conducted a 7-day A/B experiment. Table~\ref{tab:online} reports aggregated impact on (i) \textit{video volume (VV) distribution} to quantify the effectiveness on popularity debias, (ii) \textit{freshness} to measure how quickly fresh items are surfaced, (iii) \textit{engagement efficiency} to capture user-value changes (positive and negative interactions), and (iv) \textit{serving capacity} measured by Queries Per Second (QPS).

\noindent \textbf{Video View Distribution Shift.}
We bucket items by their historical view volumes (VV) and compute the relative changes of total views contributed by each bucket. A shift toward lower-VV buckets indicates that exposure is less dominated by already-popular items (i.e., reduced popularity bias). As shown in Table~\ref{tab:online}, \ToolX\ substantially increases exposure to low-/mid-popularity content ([0, 5k): +279\%; [5k, 50k): +54\%), while slightly decreasing exposure to the highest-popularity bucket ([100k, $\infty$): $-2$\%), indicating that \ToolX\ effectively redistributes traffic from head items toward the long tail.

\noindent \textbf{Freshness.}
Freshness is measured as the fraction of served items whose age (time since publish) falls within a given time window.
We report two windows: \textit{[0, 6h)} capturing ``ultra-fresh'' content discovery, and \textit{[0, 24h)} capturing same-day freshness.
\ToolX\ improves ultra-fresh exposure by +221\% and increases same-day exposure by +10\%, enabling faster delivery of newly created items.

\noindent \textbf{Engagement Efficiency.}
We track two engagement signals:
\textit{Explicit} engagement (e.g., likes, shares) and
\textit{Diversity} (i.e. interest diversity of engagement).
An increase in explicit engagement alongside an increased diversity in engaged items indicates improved engagement efficiency (higher user value per impression).
\ToolX\ yields a +0.08\% lift in explicit engagement and improves its diversity by +0.30\%, suggesting better engagement efficiency even as traffic shifts toward fresher content.

\noindent \textbf{Serving Capacity.}
We evaluate GPU server throughput using QPS (queries per second).
\ToolX\ improves serving QPS by +60\%, demonstrating substantially higher serving efficiency and increased headroom for reranking model scaling (see Appendix \ref{apx:scaling} for details).

\begin{table}[t]
\centering
\small
\caption{Aggregated 7-day online A/B testing results. \ToolX~ reduces popularity bias by shifting exposure toward low-VV items and improves freshness, engagement efficiency (more explicit and diverse engagement ), and serving throughput.}
\label{tab:online}
\vspace{-0.1in}
\begin{tabular}{p{0.6in}p{1.4in}p{0.8in}}
\hline
\textbf{Category} & \textbf{Metric} & \textbf{Relative change} \\
\hline
\multirow{2}{*}{Video view} 
  & [0, 5k)/[5k, 50k)          & +279\%/+54\%  \\
  & [50k, 100k)/[100k, $\infty$)     & +11\%/-2\%   \\
\hline
\multirow{2}{*}{Freshness}
  & [0, 6h)          & +221\% \\
  & [0, 24h)         & +10\%  \\
\hline
\multirow{2}{*}{Engagement}
  & Explicit signals & +0.08\% \\
  & Diversity  & +0.30\% \\
\hline
Throughput
  & QPS        & +60\% \\
\hline
\end{tabular}
\vspace{-3mm}
\end{table}

\section{Conclusion}
\label{sec:conclusion}
We present \ToolX, a scalable real-time retrieval framework that jointly optimizes multifaceted item embeddings and indices, eliminating the resource-intensive offline index construction and ANN search required by conventional retrieval systems. We address key production challenges through: (i) index balancing, via training-time strategies and a split-and-merge index rebalancing procedure; (ii) a unified, compact indexing structure for efficient multifaceted index lookup; and (iii) a real-time update pipeline for index refreshing. Extensive offline and online experiments on an industrial recommendation system demonstrate consistent gains in engagement efficiency, user--interest matching, and popularity debiasing, along with improved serving efficiency.


\bibliographystyle{ACM-Reference-Format}
\bibliography{reference}


\begin{thebibliography}{53}


\ifx \showCODEN    \undefined \def \showCODEN     #1{\unskip}     \fi
\ifx \showISBNx    \undefined \def \showISBNx     #1{\unskip}     \fi
\ifx \showISBNxiii \undefined \def \showISBNxiii  #1{\unskip}     \fi
\ifx \showISSN     \undefined \def \showISSN      #1{\unskip}     \fi
\ifx \showLCCN     \undefined \def \showLCCN      #1{\unskip}     \fi
\ifx \shownote     \undefined \def \shownote      #1{#1}          \fi
\ifx \showarticletitle \undefined \def \showarticletitle #1{#1}   \fi
\ifx \showURL      \undefined \def \showURL       {\relax}        \fi
\providecommand\bibfield[2]{#2}
\providecommand\bibinfo[2]{#2}
\providecommand\natexlab[1]{#1}
\providecommand\showeprint[2][]{arXiv:#2}

\bibitem[Andoni et~al\mbox{.}(2015)]%
        {andoni2015practical}
\bibfield{author}{\bibinfo{person}{Alexandr Andoni}, \bibinfo{person}{Piotr
  Indyk}, \bibinfo{person}{Thijs Laarhoven}, \bibinfo{person}{Ilya
  Razenshteyn}, {and} \bibinfo{person}{Ludwig Schmidt}.}
  \bibinfo{year}{2015}\natexlab{}.
\newblock \showarticletitle{Practical and optimal LSH for angular distance}.
\newblock \bibinfo{journal}{\emph{Advances in neural information processing
  systems}}  \bibinfo{volume}{28} (\bibinfo{year}{2015}).
\newblock


\bibitem[Aum{\"u}ller and Ceccarello(2023)]%
        {aumuller2023recent}
\bibfield{author}{\bibinfo{person}{Martin Aum{\"u}ller} {and}
  \bibinfo{person}{Matteo Ceccarello}.} \bibinfo{year}{2023}\natexlab{}.
\newblock \showarticletitle{Recent Approaches and Trends in Approximate Nearest
  Neighbor Search, with Remarks on Benchmarking.}
\newblock \bibinfo{journal}{\emph{IEEE Data Eng. Bull.}} \bibinfo{volume}{46},
  \bibinfo{number}{3} (\bibinfo{year}{2023}), \bibinfo{pages}{89--105}.
\newblock


\bibitem[Bin et~al\mbox{.}(2025)]%
        {bin2025real}
\bibfield{author}{\bibinfo{person}{Xingyan Bin}, \bibinfo{person}{Jianfei Cui},
  \bibinfo{person}{Wujie Yan}, \bibinfo{person}{Zhichen Zhao},
  \bibinfo{person}{Xintian Han}, \bibinfo{person}{Chongyang Yan},
  \bibinfo{person}{Feng Zhang}, \bibinfo{person}{Xun Zhou},
  \bibinfo{person}{Xiao Yang}, {and} \bibinfo{person}{Zuotao Liu}.}
  \bibinfo{year}{2025}\natexlab{}.
\newblock \showarticletitle{Real-time Indexing for Large-scale Recommendation
  by Streaming Vector Quantization Retriever}. In
  \bibinfo{booktitle}{\emph{Proceedings of the 31st ACM SIGKDD Conference on
  Knowledge Discovery and Data Mining V. 2}}. \bibinfo{pages}{4273--4283}.
\newblock


\bibitem[Chen et~al\mbox{.}(2024)]%
        {chen2024hllm}
\bibfield{author}{\bibinfo{person}{Junyi Chen}, \bibinfo{person}{Lu Chi},
  \bibinfo{person}{Bingyue Peng}, {and} \bibinfo{person}{Zehuan Yuan}.}
  \bibinfo{year}{2024}\natexlab{}.
\newblock \showarticletitle{HLLM: Enhancing sequential recommendations via
  hierarchical large language models for item and user modeling}.
\newblock \bibinfo{journal}{\emph{arXiv preprint arXiv:2409.12740}}
  (\bibinfo{year}{2024}).
\newblock


\bibitem[Chen et~al\mbox{.}(2022)]%
        {chen2022approximate}
\bibfield{author}{\bibinfo{person}{Rihan Chen}, \bibinfo{person}{Bin Liu},
  \bibinfo{person}{Han Zhu}, \bibinfo{person}{Yaoxuan Wang},
  \bibinfo{person}{Qi Li}, \bibinfo{person}{Buting Ma}, \bibinfo{person}{Qingbo
  Hua}, \bibinfo{person}{Jun Jiang}, \bibinfo{person}{Yunlong Xu},
  \bibinfo{person}{Hongbo Deng}, {et~al\mbox{.}}}
  \bibinfo{year}{2022}\natexlab{}.
\newblock \showarticletitle{Approximate nearest neighbor search under neural
  similarity metric for large-scale recommendation}. In
  \bibinfo{booktitle}{\emph{Proceedings of the 31st ACM International
  Conference on Information \& Knowledge Management}}.
  \bibinfo{pages}{3013--3022}.
\newblock


\bibitem[Covington et~al\mbox{.}(2016)]%
        {covington2016deep}
\bibfield{author}{\bibinfo{person}{Paul Covington}, \bibinfo{person}{Jay
  Adams}, {and} \bibinfo{person}{Emre Sargin}.}
  \bibinfo{year}{2016}\natexlab{}.
\newblock \showarticletitle{Deep neural networks for youtube recommendations}.
  In \bibinfo{booktitle}{\emph{Proceedings of the 10th ACM conference on
  recommender systems}}. \bibinfo{pages}{191--198}.
\newblock


\bibitem[Deng et~al\mbox{.}(2025)]%
        {deng2025onerec}
\bibfield{author}{\bibinfo{person}{Jiaxin Deng}, \bibinfo{person}{Shiyao Wang},
  \bibinfo{person}{Kuo Cai}, \bibinfo{person}{Lejian Ren},
  \bibinfo{person}{Qigen Hu}, \bibinfo{person}{Weifeng Ding},
  \bibinfo{person}{Qiang Luo}, {and} \bibinfo{person}{Guorui Zhou}.}
  \bibinfo{year}{2025}\natexlab{}.
\newblock \showarticletitle{Onerec: Unifying retrieve and rank with generative
  recommender and iterative preference alignment}.
\newblock \bibinfo{journal}{\emph{arXiv preprint arXiv:2502.18965}}
  (\bibinfo{year}{2025}).
\newblock


\bibitem[Douze et~al\mbox{.}(2025)]%
        {douze2025faiss}
\bibfield{author}{\bibinfo{person}{Matthijs Douze}, \bibinfo{person}{Alexandr
  Guzhva}, \bibinfo{person}{Chengqi Deng}, \bibinfo{person}{Jeff Johnson},
  \bibinfo{person}{Gergely Szilvasy}, \bibinfo{person}{Pierre-Emmanuel
  Mazar{\'e}}, \bibinfo{person}{Maria Lomeli}, \bibinfo{person}{Lucas
  Hosseini}, {and} \bibinfo{person}{Herv{\'e} J{\'e}gou}.}
  \bibinfo{year}{2025}\natexlab{}.
\newblock \showarticletitle{The faiss library}.
\newblock \bibinfo{journal}{\emph{IEEE Transactions on Big Data}}
  (\bibinfo{year}{2025}).
\newblock


\bibitem[Gao et~al\mbox{.}(2020)]%
        {gao2020deep}
\bibfield{author}{\bibinfo{person}{Weihao Gao}, \bibinfo{person}{Xiangjun Fan},
  \bibinfo{person}{Chong Wang}, \bibinfo{person}{Jiankai Sun},
  \bibinfo{person}{Kai Jia}, \bibinfo{person}{Wenzhi Xiao},
  \bibinfo{person}{Ruofan Ding}, \bibinfo{person}{Xingyan Bin},
  \bibinfo{person}{Hui Yang}, {and} \bibinfo{person}{Xiaobing Liu}.}
  \bibinfo{year}{2020}\natexlab{}.
\newblock \showarticletitle{Deep retrieval: learning a retrievable structure
  for large-scale recommendations}.
\newblock \bibinfo{journal}{\emph{arXiv preprint arXiv:2007.07203}}
  (\bibinfo{year}{2020}).
\newblock


\bibitem[Glick(1973)]%
        {glick1973sample}
\bibfield{author}{\bibinfo{person}{Ned Glick}.}
  \bibinfo{year}{1973}\natexlab{}.
\newblock \showarticletitle{Sample-based multinomial classification}.
\newblock \bibinfo{journal}{\emph{Biometrics}} (\bibinfo{year}{1973}),
  \bibinfo{pages}{241--256}.
\newblock


\bibitem[Halder et~al\mbox{.}(2024)]%
        {halder2024enhancing}
\bibfield{author}{\bibinfo{person}{Rajib~Kumar Halder},
  \bibinfo{person}{Mohammed~Nasir Uddin}, \bibinfo{person}{Md~Ashraf Uddin},
  \bibinfo{person}{Sunil Aryal}, {and} \bibinfo{person}{Ansam Khraisat}.}
  \bibinfo{year}{2024}\natexlab{}.
\newblock \showarticletitle{Enhancing K-nearest neighbor algorithm: a
  comprehensive review and performance analysis of modifications}.
\newblock \bibinfo{journal}{\emph{Journal of Big Data}} \bibinfo{volume}{11},
  \bibinfo{number}{1} (\bibinfo{year}{2024}), \bibinfo{pages}{113}.
\newblock


\bibitem[Han et~al\mbox{.}(2022)]%
        {han2022multi}
\bibfield{author}{\bibinfo{person}{Qilong Han}, \bibinfo{person}{Chi Zhang},
  \bibinfo{person}{Rui Chen}, \bibinfo{person}{Riwei Lai},
  \bibinfo{person}{Hongtao Song}, {and} \bibinfo{person}{Li Li}.}
  \bibinfo{year}{2022}\natexlab{}.
\newblock \showarticletitle{Multi-faceted global item relation learning for
  session-based recommendation}. In \bibinfo{booktitle}{\emph{Proceedings of
  the 45th international ACM SIGIR conference on research and development in
  information retrieval}}. \bibinfo{pages}{1705--1715}.
\newblock


\bibitem[Harichandan et~al\mbox{.}(2025)]%
        {harichandan2025comprehensive}
\bibfield{author}{\bibinfo{person}{Lucky Harichandan},
  \bibinfo{person}{Sasmita~Kumari Nayak}, {and} \bibinfo{person}{Satyabrata
  Lenka}.} \bibinfo{year}{2025}\natexlab{}.
\newblock \showarticletitle{A Comprehensive Review on Video Recommendation
  System: Models, Challenges, and Applications.}
\newblock \bibinfo{journal}{\emph{Journal of Engineering Science \& Technology
  Review}} \bibinfo{volume}{18}, \bibinfo{number}{3} (\bibinfo{year}{2025}).
\newblock


\bibitem[He et~al\mbox{.}(2017)]%
        {he2017neural}
\bibfield{author}{\bibinfo{person}{Xiangnan He}, \bibinfo{person}{Lizi Liao},
  \bibinfo{person}{Hanwang Zhang}, \bibinfo{person}{Liqiang Nie},
  \bibinfo{person}{Xia Hu}, {and} \bibinfo{person}{Tat-Seng Chua}.}
  \bibinfo{year}{2017}\natexlab{}.
\newblock \showarticletitle{Neural collaborative filtering}. In
  \bibinfo{booktitle}{\emph{Proceedings of the 26th international conference on
  world wide web}}. \bibinfo{pages}{173--182}.
\newblock


\bibitem[Houle and Nett(2014)]%
        {houle2014rank}
\bibfield{author}{\bibinfo{person}{Michael~E Houle} {and}
  \bibinfo{person}{Michael Nett}.} \bibinfo{year}{2014}\natexlab{}.
\newblock \showarticletitle{Rank-based similarity search: Reducing the
  dimensional dependence}.
\newblock \bibinfo{journal}{\emph{IEEE transactions on pattern analysis and
  machine intelligence}} \bibinfo{volume}{37}, \bibinfo{number}{1}
  (\bibinfo{year}{2014}), \bibinfo{pages}{136--150}.
\newblock


\bibitem[Huang et~al\mbox{.}(2025)]%
        {huang2025comprehensive}
\bibfield{author}{\bibinfo{person}{Junjie Huang}, \bibinfo{person}{Jizheng
  Chen}, \bibinfo{person}{Jianghao Lin}, \bibinfo{person}{Jiarui Qin},
  \bibinfo{person}{Ziming Feng}, \bibinfo{person}{Weinan Zhang}, {and}
  \bibinfo{person}{Yong Yu}.} \bibinfo{year}{2025}\natexlab{}.
\newblock \showarticletitle{A comprehensive survey on retrieval methods in
  recommender systems}.
\newblock \bibinfo{journal}{\emph{ACM Transactions on Information Systems}}
  \bibinfo{volume}{44}, \bibinfo{number}{1} (\bibinfo{year}{2025}),
  \bibinfo{pages}{1--43}.
\newblock


\bibitem[Hyv{\"o}nen et~al\mbox{.}(2016)]%
        {hyvonen2016fast}
\bibfield{author}{\bibinfo{person}{Ville Hyv{\"o}nen}, \bibinfo{person}{Teemu
  Pitk{\"a}nen}, \bibinfo{person}{Sotiris Tasoulis}, \bibinfo{person}{Elias
  J{\"a}{\"a}saari}, \bibinfo{person}{Risto Tuomainen}, \bibinfo{person}{Liang
  Wang}, \bibinfo{person}{Jukka Corander}, {and} \bibinfo{person}{Teemu Roos}.}
  \bibinfo{year}{2016}\natexlab{}.
\newblock \showarticletitle{Fast nearest neighbor search through sparse random
  projections and voting}. In \bibinfo{booktitle}{\emph{2016 IEEE International
  Conference on Big Data (Big Data)}}. IEEE, \bibinfo{pages}{881--888}.
\newblock


\bibitem[Jegou et~al\mbox{.}(2010)]%
        {jegou2010product}
\bibfield{author}{\bibinfo{person}{Herve Jegou}, \bibinfo{person}{Matthijs
  Douze}, {and} \bibinfo{person}{Cordelia Schmid}.}
  \bibinfo{year}{2010}\natexlab{}.
\newblock \showarticletitle{Product quantization for nearest neighbor search}.
\newblock \bibinfo{journal}{\emph{IEEE transactions on pattern analysis and
  machine intelligence}} \bibinfo{volume}{33}, \bibinfo{number}{1}
  (\bibinfo{year}{2010}), \bibinfo{pages}{117--128}.
\newblock


\bibitem[Johnson et~al\mbox{.}(2019)]%
        {johnson2019billion}
\bibfield{author}{\bibinfo{person}{Jeff Johnson}, \bibinfo{person}{Matthijs
  Douze}, {and} \bibinfo{person}{Herv{\'e} J{\'e}gou}.}
  \bibinfo{year}{2019}\natexlab{}.
\newblock \showarticletitle{Billion-scale similarity search with GPUs}.
\newblock \bibinfo{journal}{\emph{IEEE Transactions on Big Data}}
  \bibinfo{volume}{7}, \bibinfo{number}{3} (\bibinfo{year}{2019}),
  \bibinfo{pages}{535--547}.
\newblock


\bibitem[Ju et~al\mbox{.}(2025)]%
        {ju2025generative}
\bibfield{author}{\bibinfo{person}{Clark~Mingxuan Ju}, \bibinfo{person}{Liam
  Collins}, \bibinfo{person}{Leonardo Neves}, \bibinfo{person}{Bhuvesh Kumar},
  \bibinfo{person}{Louis~Yufeng Wang}, \bibinfo{person}{Tong Zhao}, {and}
  \bibinfo{person}{Neil Shah}.} \bibinfo{year}{2025}\natexlab{}.
\newblock \showarticletitle{Generative Recommendation with Semantic IDs: A
  Practitioner's Handbook}. In \bibinfo{booktitle}{\emph{Proceedings of the
  34th ACM International Conference on Information and Knowledge Management}}.
  \bibinfo{pages}{6420--6425}.
\newblock


\bibitem[Ko et~al\mbox{.}(2022)]%
        {ko2022survey}
\bibfield{author}{\bibinfo{person}{Hyeyoung Ko}, \bibinfo{person}{Suyeon Lee},
  \bibinfo{person}{Yoonseo Park}, {and} \bibinfo{person}{Anna Choi}.}
  \bibinfo{year}{2022}\natexlab{}.
\newblock \showarticletitle{A survey of recommendation systems: recommendation
  models, techniques, and application fields}.
\newblock \bibinfo{journal}{\emph{Electronics}} \bibinfo{volume}{11},
  \bibinfo{number}{1} (\bibinfo{year}{2022}), \bibinfo{pages}{141}.
\newblock


\bibitem[Lee et~al\mbox{.}(2022)]%
        {lee2022autoregressive}
\bibfield{author}{\bibinfo{person}{Doyup Lee}, \bibinfo{person}{Chiheon Kim},
  \bibinfo{person}{Saehoon Kim}, \bibinfo{person}{Minsu Cho}, {and}
  \bibinfo{person}{Wook-Shin Han}.} \bibinfo{year}{2022}\natexlab{}.
\newblock \showarticletitle{Autoregressive image generation using residual
  quantization}. In \bibinfo{booktitle}{\emph{Proceedings of the IEEE/CVF
  conference on computer vision and pattern recognition}}.
  \bibinfo{pages}{11523--11532}.
\newblock


\bibitem[Li et~al\mbox{.}(2024)]%
        {li2024full}
\bibfield{author}{\bibinfo{person}{Nian Li}, \bibinfo{person}{Yunzhu Pan},
  \bibinfo{person}{Chen Gao}, \bibinfo{person}{Depeng Jin}, {and}
  \bibinfo{person}{Qingmin Liao}.} \bibinfo{year}{2024}\natexlab{}.
\newblock \showarticletitle{Full-stage Diversified Recommendation: Large-scale
  Online Experiments in Short-video Platform}. In
  \bibinfo{booktitle}{\emph{Proceedings of the ACM Web Conference 2024}}.
  \bibinfo{pages}{4565--4574}.
\newblock


\bibitem[Luo et~al\mbox{.}(2025)]%
        {luo2025qarm}
\bibfield{author}{\bibinfo{person}{Xinchen Luo}, \bibinfo{person}{Jiangxia
  Cao}, \bibinfo{person}{Tianyu Sun}, \bibinfo{person}{Jinkai Yu},
  \bibinfo{person}{Rui Huang}, \bibinfo{person}{Wei Yuan},
  \bibinfo{person}{Hezheng Lin}, \bibinfo{person}{Yichen Zheng},
  \bibinfo{person}{Shiyao Wang}, \bibinfo{person}{Qigen Hu}, {et~al\mbox{.}}}
  \bibinfo{year}{2025}\natexlab{}.
\newblock \showarticletitle{Qarm: Quantitative alignment multi-modal
  recommendation at kuaishou}. In \bibinfo{booktitle}{\emph{Proceedings of the
  34th ACM International Conference on Information and Knowledge Management}}.
  \bibinfo{pages}{5915--5922}.
\newblock


\bibitem[Malkov et~al\mbox{.}(2014)]%
        {malkov2014approximate}
\bibfield{author}{\bibinfo{person}{Yury Malkov}, \bibinfo{person}{Alexander
  Ponomarenko}, \bibinfo{person}{Andrey Logvinov}, {and}
  \bibinfo{person}{Vladimir Krylov}.} \bibinfo{year}{2014}\natexlab{}.
\newblock \showarticletitle{Approximate nearest neighbor algorithm based on
  navigable small world graphs}.
\newblock \bibinfo{journal}{\emph{Information Systems}}  \bibinfo{volume}{45}
  (\bibinfo{year}{2014}), \bibinfo{pages}{61--68}.
\newblock


\bibitem[Malkov and Yashunin(2018)]%
        {malkov2018efficient}
\bibfield{author}{\bibinfo{person}{Yu~A Malkov} {and} \bibinfo{person}{Dmitry~A
  Yashunin}.} \bibinfo{year}{2018}\natexlab{}.
\newblock \showarticletitle{Efficient and robust approximate nearest neighbor
  search using hierarchical navigable small world graphs}.
\newblock \bibinfo{journal}{\emph{IEEE transactions on pattern analysis and
  machine intelligence}} \bibinfo{volume}{42}, \bibinfo{number}{4}
  (\bibinfo{year}{2018}), \bibinfo{pages}{824--836}.
\newblock


\bibitem[Muja and Lowe(2014)]%
        {muja2014scalable}
\bibfield{author}{\bibinfo{person}{Marius Muja} {and} \bibinfo{person}{David~G
  Lowe}.} \bibinfo{year}{2014}\natexlab{}.
\newblock \showarticletitle{Scalable nearest neighbor algorithms for high
  dimensional data}.
\newblock \bibinfo{journal}{\emph{IEEE transactions on pattern analysis and
  machine intelligence}} \bibinfo{volume}{36}, \bibinfo{number}{11}
  (\bibinfo{year}{2014}), \bibinfo{pages}{2227--2240}.
\newblock


\bibitem[Naumov et~al\mbox{.}(2019)]%
        {naumov2019deep}
\bibfield{author}{\bibinfo{person}{Maxim Naumov}, \bibinfo{person}{Dheevatsa
  Mudigere}, \bibinfo{person}{Hao-Jun~Michael Shi}, \bibinfo{person}{Jianyu
  Huang}, \bibinfo{person}{Narayanan Sundaraman}, \bibinfo{person}{Jongsoo
  Park}, \bibinfo{person}{Xiaodong Wang}, \bibinfo{person}{Udit Gupta},
  \bibinfo{person}{Carole-Jean Wu}, \bibinfo{person}{Alisson~G Azzolini},
  {et~al\mbox{.}}} \bibinfo{year}{2019}\natexlab{}.
\newblock \showarticletitle{Deep learning recommendation model for
  personalization and recommendation systems}.
\newblock \bibinfo{journal}{\emph{arXiv preprint arXiv:1906.00091}}
  (\bibinfo{year}{2019}).
\newblock


\bibitem[Paria et~al\mbox{.}(2020)]%
        {paria2020minimizing}
\bibfield{author}{\bibinfo{person}{Biswajit Paria}, \bibinfo{person}{Chih-Kuan
  Yeh}, \bibinfo{person}{Ian~EH Yen}, \bibinfo{person}{Ning Xu},
  \bibinfo{person}{Pradeep Ravikumar}, {and} \bibinfo{person}{Barnab{\'a}s
  P{\'o}czos}.} \bibinfo{year}{2020}\natexlab{}.
\newblock \showarticletitle{Minimizing flops to learn efficient sparse
  representations}.
\newblock \bibinfo{journal}{\emph{arXiv preprint arXiv:2004.05665}}
  (\bibinfo{year}{2020}).
\newblock


\bibitem[Pi et~al\mbox{.}(2020)]%
        {pi2020search}
\bibfield{author}{\bibinfo{person}{Qi Pi}, \bibinfo{person}{Guorui Zhou},
  \bibinfo{person}{Yujing Zhang}, \bibinfo{person}{Zhe Wang},
  \bibinfo{person}{Lejian Ren}, \bibinfo{person}{Ying Fan},
  \bibinfo{person}{Xiaoqiang Zhu}, {and} \bibinfo{person}{Kun Gai}.}
  \bibinfo{year}{2020}\natexlab{}.
\newblock \showarticletitle{Search-based user interest modeling with lifelong
  sequential behavior data for click-through rate prediction}. In
  \bibinfo{booktitle}{\emph{Proceedings of the 29th ACM International
  Conference on Information \& Knowledge Management}}.
  \bibinfo{pages}{2685--2692}.
\newblock


\bibitem[Rajput et~al\mbox{.}(2023)]%
        {rajput2023recommender}
\bibfield{author}{\bibinfo{person}{Shashank Rajput}, \bibinfo{person}{Nikhil
  Mehta}, \bibinfo{person}{Anima Singh}, \bibinfo{person}{Raghunandan
  Hulikal~Keshavan}, \bibinfo{person}{Trung Vu}, \bibinfo{person}{Lukasz
  Heldt}, \bibinfo{person}{Lichan Hong}, \bibinfo{person}{Yi Tay},
  \bibinfo{person}{Vinh Tran}, \bibinfo{person}{Jonah Samost}, {et~al\mbox{.}}}
  \bibinfo{year}{2023}\natexlab{}.
\newblock \showarticletitle{Recommender systems with generative retrieval}.
\newblock \bibinfo{journal}{\emph{Advances in Neural Information Processing
  Systems}}  \bibinfo{volume}{36} (\bibinfo{year}{2023}),
  \bibinfo{pages}{10299--10315}.
\newblock


\bibitem[Rangadurai et~al\mbox{.}(2024)]%
        {rangadurai2024hierarchical}
\bibfield{author}{\bibinfo{person}{Kaushik Rangadurai}, \bibinfo{person}{Siyang
  Yuan}, \bibinfo{person}{Minhui Huang}, \bibinfo{person}{Yiqun Liu},
  \bibinfo{person}{Golnaz Ghasemiesfeh}, \bibinfo{person}{Yunchen Pu},
  \bibinfo{person}{Xinfeng Xie}, \bibinfo{person}{Xingfeng He},
  \bibinfo{person}{Fangzhou Xu}, \bibinfo{person}{Andrew Cui}, {et~al\mbox{.}}}
  \bibinfo{year}{2024}\natexlab{}.
\newblock \showarticletitle{Hierarchical Structured Neural Network for
  Retrieval}.
\newblock \bibinfo{journal}{\emph{arXiv e-prints}} (\bibinfo{year}{2024}),
  \bibinfo{pages}{arXiv--2408}.
\newblock


\bibitem[Risch et~al\mbox{.}(2021)]%
        {risch2021multifaceted}
\bibfield{author}{\bibinfo{person}{Julian Risch}, \bibinfo{person}{Philipp
  Hager}, {and} \bibinfo{person}{Ralf Krestel}.}
  \bibinfo{year}{2021}\natexlab{}.
\newblock \showarticletitle{Multifaceted domain-specific document embeddings}.
  In \bibinfo{booktitle}{\emph{Proceedings of the 2021 Conference of the North
  American Chapter of the Association for Computational Linguistics: Human
  Language Technologies: Demonstrations}}. \bibinfo{pages}{78--83}.
\newblock


\bibitem[Ruder(2016)]%
        {ruder2016overview}
\bibfield{author}{\bibinfo{person}{Sebastian Ruder}.}
  \bibinfo{year}{2016}\natexlab{}.
\newblock \showarticletitle{An overview of gradient descent optimization
  algorithms}.
\newblock \bibinfo{journal}{\emph{arXiv preprint arXiv:1609.04747}}
  (\bibinfo{year}{2016}).
\newblock


\bibitem[Shrivastava and Li(2014)]%
        {shrivastava2014asymmetric}
\bibfield{author}{\bibinfo{person}{Anshumali Shrivastava} {and}
  \bibinfo{person}{Ping Li}.} \bibinfo{year}{2014}\natexlab{}.
\newblock \showarticletitle{Asymmetric LSH (ALSH) for sublinear time maximum
  inner product search (MIPS)}.
\newblock \bibinfo{journal}{\emph{Advances in neural information processing
  systems}}  \bibinfo{volume}{27} (\bibinfo{year}{2014}).
\newblock


\bibitem[Simhadri et~al\mbox{.}(2022)]%
        {simhadri2022results}
\bibfield{author}{\bibinfo{person}{Harsha~Vardhan Simhadri},
  \bibinfo{person}{George Williams}, \bibinfo{person}{Martin Aum{\"u}ller},
  \bibinfo{person}{Matthijs Douze}, \bibinfo{person}{Artem Babenko},
  \bibinfo{person}{Dmitry Baranchuk}, \bibinfo{person}{Qi Chen},
  \bibinfo{person}{Lucas Hosseini}, \bibinfo{person}{Ravishankar
  Krishnaswamny}, \bibinfo{person}{Gopal Srinivasa}, {et~al\mbox{.}}}
  \bibinfo{year}{2022}\natexlab{}.
\newblock \showarticletitle{Results of the NeurIPS’21 challenge on
  billion-scale approximate nearest neighbor search}. In
  \bibinfo{booktitle}{\emph{NeurIPS 2021 Competitions and Demonstrations
  Track}}. PMLR, \bibinfo{pages}{177--189}.
\newblock


\bibitem[Singh et~al\mbox{.}(2024)]%
        {singh2024better}
\bibfield{author}{\bibinfo{person}{Anima Singh}, \bibinfo{person}{Trung Vu},
  \bibinfo{person}{Nikhil Mehta}, \bibinfo{person}{Raghunandan Keshavan},
  \bibinfo{person}{Maheswaran Sathiamoorthy}, \bibinfo{person}{Yilin Zheng},
  \bibinfo{person}{Lichan Hong}, \bibinfo{person}{Lukasz Heldt},
  \bibinfo{person}{Li Wei}, \bibinfo{person}{Devansh Tandon}, {et~al\mbox{.}}}
  \bibinfo{year}{2024}\natexlab{}.
\newblock \showarticletitle{Better generalization with semantic ids: A case
  study in ranking for recommendations}. In
  \bibinfo{booktitle}{\emph{Proceedings of the 18th ACM Conference on
  Recommender Systems}}. \bibinfo{pages}{1039--1044}.
\newblock


\bibitem[Spring and Shrivastava(2017)]%
        {spring2017new}
\bibfield{author}{\bibinfo{person}{Ryan Spring} {and}
  \bibinfo{person}{Anshumali Shrivastava}.} \bibinfo{year}{2017}\natexlab{}.
\newblock \showarticletitle{A new unbiased and efficient class of lsh-based
  samplers and estimators for partition function computation in log-linear
  models}.
\newblock \bibinfo{journal}{\emph{arXiv preprint arXiv:1703.05160}}
  (\bibinfo{year}{2017}).
\newblock


\bibitem[Van Den~Oord et~al\mbox{.}(2017)]%
        {van2017neural}
\bibfield{author}{\bibinfo{person}{Aaron Van Den~Oord}, \bibinfo{person}{Oriol
  Vinyals}, {et~al\mbox{.}}} \bibinfo{year}{2017}\natexlab{}.
\newblock \showarticletitle{Neural discrete representation learning}.
\newblock \bibinfo{journal}{\emph{Advances in neural information processing
  systems}}  \bibinfo{volume}{30} (\bibinfo{year}{2017}).
\newblock


\bibitem[Wu et~al\mbox{.}(2024)]%
        {wu2024effectiveness}
\bibfield{author}{\bibinfo{person}{Jiancan Wu}, \bibinfo{person}{Xiang Wang},
  \bibinfo{person}{Xingyu Gao}, \bibinfo{person}{Jiawei Chen},
  \bibinfo{person}{Hongcheng Fu}, {and} \bibinfo{person}{Tianyu Qiu}.}
  \bibinfo{year}{2024}\natexlab{}.
\newblock \showarticletitle{On the effectiveness of sampled softmax loss for
  item recommendation}.
\newblock \bibinfo{journal}{\emph{ACM Transactions on Information Systems}}
  \bibinfo{volume}{42}, \bibinfo{number}{4} (\bibinfo{year}{2024}),
  \bibinfo{pages}{1--26}.
\newblock


\bibitem[Xu et~al\mbox{.}(2025)]%
        {xu2025mmq}
\bibfield{author}{\bibinfo{person}{Yi Xu}, \bibinfo{person}{Moyu Zhang},
  \bibinfo{person}{Chaofan Fan}, \bibinfo{person}{Jinxin Hu},
  \bibinfo{person}{Xiaochen Li}, \bibinfo{person}{Yu Zhang},
  \bibinfo{person}{Xiaoyi Zeng}, {and} \bibinfo{person}{Jing Zhang}.}
  \bibinfo{year}{2025}\natexlab{}.
\newblock \showarticletitle{MMQ-v2: Align, Denoise, and Amplify: Adaptive
  Behavior Mining for Semantic IDs Learning in Recommendation}.
\newblock \bibinfo{journal}{\emph{arXiv preprint arXiv:2510.25622}}
  (\bibinfo{year}{2025}).
\newblock


\bibitem[Xue et~al\mbox{.}(2025)]%
        {xue2025silvertorch}
\bibfield{author}{\bibinfo{person}{Bi Xue}, \bibinfo{person}{Hong Wu},
  \bibinfo{person}{Lei Chen}, \bibinfo{person}{Chao Yang},
  \bibinfo{person}{Yiming Ma}, \bibinfo{person}{Fei Ding},
  \bibinfo{person}{Zhen Wang}, \bibinfo{person}{Liang Wang},
  \bibinfo{person}{Xiaoheng Mao}, \bibinfo{person}{Ke Huang}, {et~al\mbox{.}}}
  \bibinfo{year}{2025}\natexlab{}.
\newblock \showarticletitle{SilverTorch: A Unified Model-based System to
  Democratize Large-Scale Recommendation on GPUs}.
\newblock \bibinfo{journal}{\emph{arXiv preprint arXiv:2511.14881}}
  (\bibinfo{year}{2025}).
\newblock


\bibitem[Yan et~al\mbox{.}(2026)]%
        {yan2026merge}
\bibfield{author}{\bibinfo{person}{Jing Yan}, \bibinfo{person}{Yimeng Bai},
  \bibinfo{person}{Zongyu Liu}, \bibinfo{person}{Yahui Liu},
  \bibinfo{person}{Junwei Wang}, \bibinfo{person}{Jingze Huang},
  \bibinfo{person}{Haoda Li}, \bibinfo{person}{Sihao Ding},
  \bibinfo{person}{Shaohui Ruan}, {and} \bibinfo{person}{Yang Zhang}.}
  \bibinfo{year}{2026}\natexlab{}.
\newblock \showarticletitle{MERGE: Next-Generation Item Indexing Paradigm for
  Large-Scale Streaming Recommendation}.
\newblock \bibinfo{journal}{\emph{arXiv preprint arXiv:2601.20199}}
  (\bibinfo{year}{2026}).
\newblock


\bibitem[Yan et~al\mbox{.}(2024)]%
        {yan2024trinity}
\bibfield{author}{\bibinfo{person}{Jing Yan}, \bibinfo{person}{Liu Jiang},
  \bibinfo{person}{Jianfei Cui}, \bibinfo{person}{Zhichen Zhao},
  \bibinfo{person}{Xingyan Bin}, \bibinfo{person}{Feng Zhang}, {and}
  \bibinfo{person}{Zuotao Liu}.} \bibinfo{year}{2024}\natexlab{}.
\newblock \showarticletitle{Trinity: Syncretizing Multi-/Long-Tail/Long-Term
  Interests All in One}. In \bibinfo{booktitle}{\emph{Proceedings of the 30th
  ACM SIGKDD Conference on Knowledge Discovery and Data Mining}}.
  \bibinfo{pages}{6095--6104}.
\newblock


\bibitem[Yang et~al\mbox{.}(2023)]%
        {yang2023generic}
\bibfield{author}{\bibinfo{person}{Zhengyi Yang}, \bibinfo{person}{Xiangnan
  He}, \bibinfo{person}{Jizhi Zhang}, \bibinfo{person}{Jiancan Wu},
  \bibinfo{person}{Xin Xin}, \bibinfo{person}{Jiawei Chen}, {and}
  \bibinfo{person}{Xiang Wang}.} \bibinfo{year}{2023}\natexlab{}.
\newblock \showarticletitle{A generic learning framework for sequential
  recommendation with distribution shifts}. In
  \bibinfo{booktitle}{\emph{Proceedings of the 46th International ACM SIGIR
  Conference on Research and Development in Information Retrieval}}.
  \bibinfo{pages}{331--340}.
\newblock


\bibitem[Zhai et~al\mbox{.}(2023)]%
        {zhai2023revisiting}
\bibfield{author}{\bibinfo{person}{Jiaqi Zhai}, \bibinfo{person}{Zhaojie Gong},
  \bibinfo{person}{Yueming Wang}, \bibinfo{person}{Xiao Sun},
  \bibinfo{person}{Zheng Yan}, \bibinfo{person}{Fu Li}, {and}
  \bibinfo{person}{Xing Liu}.} \bibinfo{year}{2023}\natexlab{}.
\newblock \showarticletitle{Revisiting Neural Retrieval on Accelerators}. In
  \bibinfo{booktitle}{\emph{Proceedings of the 29th ACM SIGKDD Conference on
  Knowledge Discovery and Data Mining}}. \bibinfo{pages}{5520--5531}.
\newblock


\bibitem[Zhai et~al\mbox{.}(2024)]%
        {zhai2024actions}
\bibfield{author}{\bibinfo{person}{Jiaqi Zhai}, \bibinfo{person}{Lucy Liao},
  \bibinfo{person}{Xing Liu}, \bibinfo{person}{Yueming Wang},
  \bibinfo{person}{Rui Li}, \bibinfo{person}{Xuan Cao}, \bibinfo{person}{Leon
  Gao}, \bibinfo{person}{Zhaojie Gong}, \bibinfo{person}{Fangda Gu},
  \bibinfo{person}{Michael He}, {et~al\mbox{.}}}
  \bibinfo{year}{2024}\natexlab{}.
\newblock \showarticletitle{Actions speak louder than words: Trillion-parameter
  sequential transducers for generative recommendations}.
\newblock \bibinfo{journal}{\emph{arXiv preprint arXiv:2402.17152}}
  (\bibinfo{year}{2024}).
\newblock


\bibitem[Zhang et~al\mbox{.}(2024)]%
        {zhang2024robust}
\bibfield{author}{\bibinfo{person}{An Zhang}, \bibinfo{person}{Wenchang Ma},
  \bibinfo{person}{Jingnan Zheng}, \bibinfo{person}{Xiang Wang}, {and}
  \bibinfo{person}{Tat-Seng Chua}.} \bibinfo{year}{2024}\natexlab{}.
\newblock \showarticletitle{Robust collaborative filtering to popularity
  distribution shift}.
\newblock \bibinfo{journal}{\emph{ACM Transactions on Information Systems}}
  \bibinfo{volume}{42}, \bibinfo{number}{3} (\bibinfo{year}{2024}),
  \bibinfo{pages}{1--25}.
\newblock


\bibitem[Zhang et~al\mbox{.}(2025)]%
        {zhang2025optimizing}
\bibfield{author}{\bibinfo{person}{Jiang Zhang}, \bibinfo{person}{Sumit Kumar},
  \bibinfo{person}{Wei Chang}, \bibinfo{person}{Yubo Wang},
  \bibinfo{person}{Feng Zhang}, \bibinfo{person}{Weize Mao},
  \bibinfo{person}{Hanchao Yu}, \bibinfo{person}{Aashu Singh},
  \bibinfo{person}{Min Li}, {and} \bibinfo{person}{Qifan Wang}.}
  \bibinfo{year}{2025}\natexlab{}.
\newblock \showarticletitle{Optimizing Recall or Relevance? A Multi-Task
  Multi-Head Approach for Item-to-Item Retrieval in Recommendation}. In
  \bibinfo{booktitle}{\emph{Proceedings of the 31st ACM SIGKDD Conference on
  Knowledge Discovery and Data Mining V. 2}}. \bibinfo{pages}{5194--5204}.
\newblock


\bibitem[Zhang et~al\mbox{.}(2022)]%
        {zhang2022deep}
\bibfield{author}{\bibinfo{person}{Jihai Zhang}, \bibinfo{person}{Fangquan
  Lin}, \bibinfo{person}{Cheng Yang}, {and} \bibinfo{person}{Wei Wang}.}
  \bibinfo{year}{2022}\natexlab{}.
\newblock \showarticletitle{Deep multi-representational item network for CTR
  prediction}. In \bibinfo{booktitle}{\emph{Proceedings of the 45th
  International ACM SIGIR Conference on Research and Development in Information
  Retrieval}}. \bibinfo{pages}{2277--2281}.
\newblock


\bibitem[Zhou et~al\mbox{.}(2025)]%
        {zhou2025onerec}
\bibfield{author}{\bibinfo{person}{Guorui Zhou}, \bibinfo{person}{Hengrui Hu},
  \bibinfo{person}{Hongtao Cheng}, \bibinfo{person}{Huanjie Wang},
  \bibinfo{person}{Jiaxin Deng}, \bibinfo{person}{Jinghao Zhang},
  \bibinfo{person}{Kuo Cai}, \bibinfo{person}{Lejian Ren}, \bibinfo{person}{Lu
  Ren}, \bibinfo{person}{Liao Yu}, {et~al\mbox{.}}}
  \bibinfo{year}{2025}\natexlab{}.
\newblock \showarticletitle{Onerec-v2 technical report}.
\newblock \bibinfo{journal}{\emph{arXiv preprint arXiv:2508.20900}}
  (\bibinfo{year}{2025}).
\newblock


\bibitem[Zhu et~al\mbox{.}(2019)]%
        {zhu2019joint}
\bibfield{author}{\bibinfo{person}{Han Zhu}, \bibinfo{person}{Daqing Chang},
  \bibinfo{person}{Ziru Xu}, \bibinfo{person}{Pengye Zhang},
  \bibinfo{person}{Xiang Li}, \bibinfo{person}{Jie He}, \bibinfo{person}{Han
  Li}, \bibinfo{person}{Jian Xu}, {and} \bibinfo{person}{Kun Gai}.}
  \bibinfo{year}{2019}\natexlab{}.
\newblock \showarticletitle{Joint optimization of tree-based index and deep
  model for recommender systems}.
\newblock \bibinfo{journal}{\emph{Advances in Neural Information Processing
  Systems}}  \bibinfo{volume}{32} (\bibinfo{year}{2019}).
\newblock


\bibitem[Zhu et~al\mbox{.}(2018)]%
        {zhu2018learning}
\bibfield{author}{\bibinfo{person}{Han Zhu}, \bibinfo{person}{Xiang Li},
  \bibinfo{person}{Pengye Zhang}, \bibinfo{person}{Guozheng Li},
  \bibinfo{person}{Jie He}, \bibinfo{person}{Han Li}, {and}
  \bibinfo{person}{Kun Gai}.} \bibinfo{year}{2018}\natexlab{}.
\newblock \showarticletitle{Learning tree-based deep model for recommender
  systems}. In \bibinfo{booktitle}{\emph{Proceedings of the 24th ACM SIGKDD
  international conference on knowledge discovery \& data mining}}.
  \bibinfo{pages}{1079--1088}.
\newblock


\end{thebibliography}

\appendix

\begin{table*}[t]
\centering
\small
\caption{Effects of number of facets on \ToolX's performance. Note that VVC, Like, LWT, CCD represent video view complete, like, long watch time, cold content delivery tasks respectively.}
\vspace{-2mm}
\label{tab:facet_full}
\begin{tabular}{p{0.75in}|p{0.75in}p{0.75in}p{0.75in}p{0.75in}|p{0.75in}p{0.75in}|p{0.6in}}
\toprule
\multirow{2}{*}{Num. of facets} & \multicolumn{4}{c}{Recall} & \multicolumn{2}{c}{Semantic relevance} & \multirow{2}{*}{Throughput}\\ \cline{2-7}
 & VVC & Like & LWT & CCD & $I_1$ topic & $I_2$ topic & \\
\midrule

One      & 20.72\% (-13.81\%)        & 24.19\% (-9.47\%)
         & 24.49\% (-10.91\%)        & 10.53\% (-11.21\%)
         & 45.52\% (-12.98\%)        & 31.51\% (-18.58\%) & +0.0\%\\

\underline{Two (default)}             & \underline{24.04\%}        & \underline{26.72\%}
                                      & \underline{27.49\%}        & \underline{11.86\%}
                                      & \underline{52.31\%}        & \underline{38.70\%} & -18.8\% \\

Three    & 24.13\% (+0.37\%)        & 27.53\% (+3.03\%)
         & 28.19\% (+2.55\%)        & 12.23\% (+3.12\%)
         & 50.60\% (-3.27\%)        & 37.11\% (-4.11\%)  & -23.9\% \\
\bottomrule
\end{tabular}
\end{table*}

\begin{table*}[t]
\centering
\small
\caption{Effects of codebook size on \ToolX's performance. Note that VVC, Like, LWT, CCD represent video view complete, like, long watch time, cold content delivery tasks respectively.}
\vspace{-2mm}
\label{tab:codebook_full}
\begin{tabular}{p{0.9in}|p{0.8in}p{0.8in}p{0.8in}p{0.85in}|p{0.9in}p{0.9in}}
\toprule
\multirow{2}{*}{Codebook size} & \multicolumn{4}{c}{Recall} & \multicolumn{2}{c}{Semantic relevance}\\ \cline{2-7}
 & VVC & Like & LWT & CCD & $I_1 $ topic & $I_2$ topic \\
\midrule
10k     & 22.54\% (-6.24\%)        & 25.51\% (-4.53\%)
        & 26.71\% (-2.84\%)        & 10.82\% (-8.77\%)
        & 55.10\% (+5.33\%)        & 43.10\% (+11.37\%) \\

256$\times$128           & 24.07\% (+0.12\%)        & 26.82\% (+0.37\%)
                         & 27.53\% (+0.15\%)        & 11.43\% (-3.63\%)
                         & 52.21\% (-0.19\%)        & 38.76\% (+0.16\%) \\

\underline{512$\times$128 (default)} & \underline{24.04\%}        & \underline{26.72\%}
                                     & \underline{27.49\%}        & \underline{11.86\%}
                                     & \underline{52.31\%}        & \underline{38.70\%} \\

1024$\times$256           & 23.18\% (-3.58\%)        & 25.97\% (-2.81\%)
                          & 27.15\% (-1.24\%)        & 11.98\% (+1.01\%)
                          & 52.73\% (+0.80\%)        & 39.49\% (+2.04\%) \\
\bottomrule
\end{tabular}
\end{table*}

\section{Training Details}
\label{apx:param}
\begin{table}[h]
\caption{Hyperparameters of \ToolX.}
\vspace{-2mm}
\label{tab:model_hyperparameters}
\centering
\small
\begin{tabular}{p{1.5in}|p{1.5in}}
\toprule
\centering
\textbf{Name} & \textbf{Value} \\
\hline
Number of GPUs & 48 A100s \\
Batch size & 2048 \\
Learning rate & 0.01 \\
Optimizer & Adagrad \\
Training epoch & 1 \\
Item embedding dim & 128 \\
Codeword embedding dim & 128 \\
\bottomrule
\end{tabular}
\end{table}

\noindent\textbf{Training setups.} We train \ToolX\ alongside the baseline models on real data from Period~$P_1$, and evaluate all models on real data from Period~$P_2$, which follows $P_1$. For \ToolX, we set the total number of selected indices to $K=200$ and the number of items retained after reranking to $N=15$. For offline evaluation, top-k index selection is enabled. To compute recall@1000, we sort retrieved items by their reranking scores and keep the top 1000. We use 1{,}000 because, in industrial systems, the retrieval stage typically passes on the order of 1k candidates to downstream stages \cite{covington2016deep}. We train the model on 48 A100 GPUs with a batch size of 2048 per GPU. We use the Adagrad optimizer with a learning rate of 0.01, and each training example is used exactly once. Both item embeddings and codewords have dimension 128 (see Table~\ref{tab:model_hyperparameters} for details).

\noindent\textbf{Codebook regularization loss.} As mentioned in Section~\ref{subsec:train}, we introduce a \textit{regularization term} that penalizes over-utilized (i.e., overly popular) codewords to encourage balanced codeword usage, following~\cite{rangadurai2024hierarchical}. Specifically, for each codebook layer $C_l \in \mathbb{R}^{F \times N_l \times d}$ and a batch of item (residual) embeddings $r_{l-1} \in \mathbb{R}^{F \times B \times d}$, we define the regularization loss as:
\begin{equation}
    Loss(C_l, r_{l-1})=\frac{1}{FN_l}\sum_{f,j}(\frac{\sum_{k}||C_l[f,j]-r_{l-1}[f,k]||_2}{B})^2.
\end{equation}
It is used to minimize the sum of squares of the mean
distance between each codeword and all item embeddings in a batch. This loss is originally introduced in \cite{paria2020minimizing} to ensure that the in-batch cluster distribution is even.

\section{Index Selection Strategies}
\label{apx:index}
After receiving $T \times F$ indices from the index lookup module, the index selection module selects $K$ unique indices. Each selected index represents an \emph{interest bucket} containing a set of correlated items that a user may be interested in. Each dimension $\tilde{c}_f(t)$ corresponds to a single-facet item index for facet $f \in \{1,\ldots,F\}$. Because multiple trigger items may map to the same single-facet index, we aggregate these mappings into an index--frequency histogram for each facet, which can be used for multi-interest mining~\cite{yan2024trinity}. Specifically, for facet $f$ we compute:
\begin{equation}
\label{eq:serving:hist}
h_f(m) \;=\; \sum_{t \in \mathcal{T}} \mathbf{1}\!\left[\tilde{c}_f(t) = m\right], 
\qquad m \in \{0,\ldots,M_f-1\},
\end{equation}
and its normalized distribution
\begin{equation}
\label{eq:serving:dist}
p_f(m) \;=\; \frac{h_f(m)}{\sum_{m'} h_f(m')}.
\end{equation}
Here, $\mathcal{T}$ denotes the set of trigger items selected from the user's interaction history; $h_f(m)$ counts how many triggers map to index $m$ under facet $f$; and $p_f(m)$ estimates the selection probability of index $m$. Intuitively, higher-frequency indices provide stronger evidence of user preference for the corresponding interest bucket. With this, \ToolX~  applies user-interest mining approaches similar to~\cite{yan2024trinity} and runs several index selection strategies in parallel. We denote the selected indices for facet $f$ by $\mathcal{S}_f$, and the final merged set by
\[
\mathcal{S} \;=\; \bigcup_{f=1}^{F} \mathcal{S}_f,
\]
typically with $|\mathcal{S}| = O(10^2)$.Below we summarize several index selection strategies used in \ToolX:

\noindent\textbf{Multi-interest extraction.} \ToolX~ applies multinomial sampling over the index-frequency distribution to select top-$k$ indices,  so that indices supported by more trigger items are more likely to be selected. Specifically, we sample from $p_f$ with a temperature $\tau>0$:
\begin{equation}
\label{eq:serving:temp}
\pi_f(m) \;=\; \frac{p_f(m)^{1/\tau}}{\sum_{m'} p_f(m')^{1/\tau}},
\end{equation}
and draw $k_1$ indices from $\pi_f$ to form $\mathcal{S}_f$. This encourages the selected indices to match the user's multi-interest profile while allowing controllable exploration via $\tau$. By default, we use $\tau=1.0$ 

\noindent\textbf{Recent-interest boosting.} Since the trigger set $\mathcal{T}$ is time-ordered, for indices associated with the most recent triggers, \ToolX~ increases their priority to capture users’ latest interests more responsively. In practice, this can be implemented by upweighting histogram counts contributed by recent triggers (equivalently, modifying $h_f$ before normalization), or by injecting a small set of indices derived from the most recent triggers directly into $\mathcal{S}_f$.
    
\noindent\textbf{Long-tail interest exploration.} For long-tail users with limited interaction history, fewer trigger items are extracted, which yields fewer mapped indices. To mitigate this, for each selected index, we additionally include neighboring indices that share the same upper-level indices, enabling more diverse retrieval for exploration.

\noindent\textbf{Retrieve more/less for strong/weak interests.} Higher/lower index frequency implies stronger/weaker interest signals. To reflect this, \ToolX~ passes $h_f(m)$ to \textit{Per-index rerank} module, in order to control how many items are selected per index. For example, given a facet-level candidate item quota $Q_f^{\mathrm{tot}}$, we allocate an index-level item quota as:
\begin{equation}
\label{eq:serving:quota}
Q_f(m) \;=\; \left\lceil
Q_f^{\mathrm{tot}}\cdot \frac{\left(h_f(m)\right)^{\alpha}}{\sum_{m'\in\mathcal{S}_f}\left(h_f(m')\right)^{\alpha}}
\right\rceil,
\end{equation}
where $\alpha\ge 0$ controls how aggressively we prioritize strong interests (with $\alpha=0$ yielding uniform allocation).

These selection paths operate in parallel for each facet, and we merge the indices selected by all paths across all facets to obtain the final set $\mathcal{S}$. Notably, if we fix the total number of selected indices $|\mathcal{S}|$, increasing the number of facets does not increase the reranking cost; therefore, the serving overhead remains bounded.

\section{Scaling \ToolX}
\label{apx:scaling}
There are two ways to scale \ToolX. The first is to increase the number of facets, where each facet is trained with a distinct objective so that multi-way retrieval can be performed at serving time. Note that, by controlling the total number of selected indices, this does not increase serving cost linearly. 
Second, since \ToolX{} does not rely on ANN search, its end-to-end throughput is significantly higher than ANN-based approaches. This leaves headroom to use more complicated reranker models, improving reranking precision.

\section{Supplementary Evaluation Results}
\label{apx:full}
In this section, we study the sensitivity of \ToolX{} to two key hyperparameters: the \textit{number of facets} and the \textit{codebook size}. We report recall on four engagement tasks and item semantic relevance on $T_1$/$T_2$ topics under different settings in Tables~\ref{tab:facet_full} and~\ref{tab:codebook_full}, respectively.

\noindent \textbf{Number of Facets.} We vary the number of facets from 1 to 3. The first facet is trained on an aggregated engagement objective (i.e., the sum of engagement labels). The second facet emphasizes item semantic relevance via multi-task learning, following \cite{zhang2025optimizing}. The third facet is specialized for time-spent-related engagement signals.
For a fair comparison, we fix the total number of retrieved items by adjusting the number of selected indices per facet. Specifically, for the single-facet model we retrieve the top-200 indices. For the two-facet model, we retrieve 100 indices per facet. For the three-facet model, we retrieve 75/50/75 indices from facets 1/2/3, respectively.

As shown in Table~\ref{tab:facet_full}, increasing the number of facets (from 1 to 3) consistently improves recall because each facet’s item embeddings and indices are optimized for a distinct objective—aggregated engagement, semantic relevance, or time spent—so different facets retrieve complementary sets of correlated items. Moreover, because we keep the total number of selected indices fixed across settings, the downstream reranking cost remains unchanged; as a result, end-to-end serving overhead does not scale linearly with the number of facets.

\noindent \textbf{Codebook Size.}
We compare four codebook variants: a single-layer codebook of size 10k and three two-layer codebooks with increasing capacity. As shown in Table~\ref{tab:codebook_full}, the single-layer 10k setting consistently regresses recall across all tasks (e.g., VVC: $-6.24\%$, Like: $-4.53\%$, LWT: $-2.84\%$, CCD: $-8.77\%$) despite achieving the strongest interest matching (e.g., $I_1$ topic: $+5.33\%$, $I_2$ topic: $+11.37\%$), suggesting that coarse quantization yields semantically coherent but less discriminative clusters. Among the two-layer codebooks, 256$\times$128 and 512$\times$128 deliver the best overall recall, while larger capacity (1024$\times$256) improves interest matching ($I_1$: $+0.80\%$, $I_2$: $+2.04\%$) at the cost of lower recall (e.g., VVC: $-3.58\%$, LWT: $-1.24\%$). Overall, 512$\times$128 provides a robust balance between recall and interest matching and is used as the default setting.

\end{document}